\documentclass[sigconf, nonacm]{acmart}

\usepackage{makecell}
\usepackage{multirow}
\usepackage[table,xcdraw]{xcolor}
\usepackage{booktabs}
\usepackage{graphicx}
\usepackage{caption}
\usepackage{subcaption}
\usepackage{dblfloatfix}
\usepackage{url}
\AtBeginDocument{%
  }

\begin{document}

\title{TileFuse: A Fused Mixed-Precision Kernel Library for Efficient Quantized LLM Inference on AMD NPUs}

\author{Wesley Pang}
\authornote{Both authors contributed equally to this research.}
\affiliation{%
  \institution{University of Illinois Urbana-Champaign}
  \department{Department of Electrical and Computer Engineering}
  \country{USA}
}

\author{Gregory Hyegang Jun}
\authornotemark[1]
\affiliation{%
  \institution{University of Illinois Urbana-Champaign}
  \department{Department of Electrical and Computer Engineering}
  \country{USA}
}

\author{Feiyang Liu}
\affiliation{%
  \institution{University of Illinois Urbana-Champaign}
  \department{Department of Electrical and Computer Engineering}
  \country{USA}
}

\author{Deming Chen}
\affiliation{%
  \institution{University of Illinois Urbana-Champaign}
  \department{Department of Electrical and Computer Engineering}
  \country{USA}
}

\begin{abstract}
With the growing demand for on-device LLM inference, edge SoCs increasingly integrate NPUs to improve performance and energy efficiency under tight power and thermal budgets. However, practical LLM deployment on current client NPUs remains difficult: widely used quantization formats such as AWQ do not map cleanly onto many existing NPU software stacks, which are often proprietary and expose limited low-level control. In this work, we present \textit{TileFuse}, a close-to-metal mixed-precision kernel library for AMD XDNA2 NPUs that targets GEMM/GEMV-based operators in quantized LLM inference. TileFuse brings practical low-bit formats such as AWQ-style $W4A16$ and $W8A16$ directly onto XDNA2, rather than forcing the model to be reshaped around an NPU-specific quantization scheme. 

TileFuse co-designs weight layout, metadata placement, mixed-precision microkernels, and array-level dataflow. Specifically, it fuses unpacking, dequantization, and GEMM/GEMV execution into a single kernel flow, introduces an interleaved pre-tiling layout that supports GEMM dimensions up to 32K, and redesigns GEMV dataflow to utilize the full $4\times8$ AIE array. Across kernel-level evaluations, TileFuse improves performance by up to 121.6\% for GEMM and 281\% for GEMV over full-precision baselines, while delivering more than $2\times$ performance and energy-efficiency gains over strong iGPU baselines on GEMM. In end-to-end LLM experiments on Ryzen AI laptops, TileFuse achieves up to $2.0\times$ lower prefilling latency with more than 64.6\% lower energy consumption. Together, these results show that XDNA2 is a practical target for AWQ-style edge LLM inference and that native NPU support for off-the-shelf quantization can make NPUs substantially more usable in real client deployments.\footnote{Source code: https://github.com/glassescrab/mlir-aie/tree/feature/update-mix-mm-int4-verification}

\end{abstract}



\maketitle
    
\section{Introduction}
\label{sec:introduction}

\begin{figure}[t]
  \centering
  \includegraphics[width=1\linewidth]{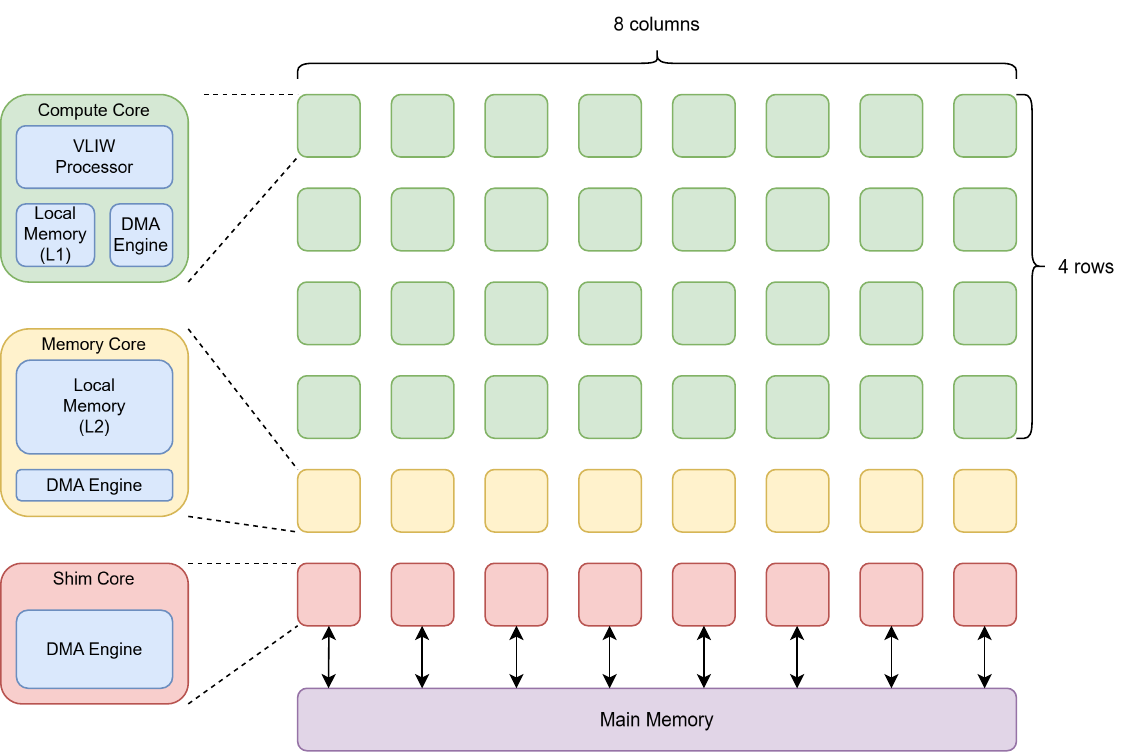}
  \caption{AMD XDNA2 NPU's spatial dataflow architecture.}
  \label{fig:AMD_XDNA}
\end{figure}

Transformer-based models such as large language models, vision transformers, vision-language-action models, and many others~\cite{siriwardhana2021survey,tahir2025edge,openvla_2024,pi0_2024,hariharan2023realtime,wang2024mememo,wang2025leann,wang2025empowering} are fundamentally transforming how we interact with machines and data, while also raising the compute demands of modern systems~\cite{openai_chatgpt_2023,openai_gpt4_2023,anthropic_claude_code_2024,openai_khan_academy_2023}. At the same time, advances in model quantization and distillation are bringing increasingly capable models to the edge~\cite{google_gemma4_model_card_2026}. To meet these growing edge-AI demands, modern system-on-chips (SoCs) have increasingly embraced specialization by integrating neural processing units (NPUs) as a core component alongside CPUs and iGPUs. This specialization is motivated by the need to deliver higher performance and better energy efficiency for AI workloads within the tight power, thermal, and area constraints of edge devices~\cite{amdnpu,intelnpu,qualcommnpu,applenpu}.

Despite this promise, NPUs remain difficult to use as first-class LLM accelerators. The main obstacle is not simply raw compute capability, but fragmentation across architectures, software stacks, and programming models. In practice, many NPU frameworks remain proprietary and expose only high-level deployment interfaces rather than low-level programming control. In AMD’s case, NPU execution is exposed only for select models through ONNX Runtime~\cite{amd_oga_2026}; Apple exposes the NPU only through Core ML~\cite{apple_coreml_2026} as a backend target; Intel primarily through OpenVINO~\cite{intel_openvino_npu_2025,openvino_npu_device_2025}; and Qualcomm through QNN / AI Engine Direct~\cite{qualcomm_qnn_2026}. As a result, unlike GPUs, NPUs still lack a broadly portable low-level software substrate on which developers can easily build custom kernels, runtime policies, and cross-accelerator scheduling mechanisms. Practical quantization formats such as AWQ~\cite{lin2023awq} are widely used and distributed because they substantially reduce memory footprint with minimal impact on model quality, yet NPU support for such formats remains limited, largely because of proprietary stacks and the lack of usable low-level programming frameworks.

In this context, NPUs that do expose differing levels of low-level programmability have shown promise. On Qualcomm, ScalingLLM~\cite{scaling_llm_mobile_2025} demonstrates a hardware-aware block-wise quantization scheme that strikes a balance between NPU architectural constraints and LLM quantization requirements, allowing the model to retain strong accuracy while achieving good performance. On Ryzen NPUs, Wang et al.~\cite{wang2025asymmetrictilebufferingbeneficial} show that, by leveraging the open low-level programming framework IRON~\cite{hunhoff2025efficiencyexpressivityextensibilityclosetometal}, block-floating-point~\cite{song2018computation} general matrix multiply (GEMM) kernels can approach up to 28 TOP/s.

However, both directions still fall short in ways that limit broader adoption by the LLM community. In the case of ScalingLLM~\cite{scaling_llm_mobile_2025}, the system does not implement standard AWQ directly, but instead relies on an NPU-tailored block-wise quantization scheme to accommodate Qualcomm’s NPU architecture. In the case of Wang et al.~\cite{wang2025asymmetrictilebufferingbeneficial}, although the reported GEMM throughput is strong and the use of IRON is promising, the kernels cannot be used for practical LLM deployment: the supported GEMM dimensions do not fully cover the large FFN shapes common in models such as Gemma-2B~\cite{gemma}, and the reduced precision of the BFP16 kernels is not sufficient for accurate end-to-end LLM inference. As a result, across both current SoC platforms and current academic work, NPUs remain difficult for the broader LLM community to adopt. Users are often forced to adapt the model to the NPU through proprietary quantization schemes or specialized data formats, rather than being able to run off-the-shelf pre-quantized models directly on the hardware.

Against this backdrop, AMD Ryzen’s XDNA2~\cite{amdnpu, rico2024xdna} offers a unique combination of capability and programmability: a reconfigurable spatial architecture together with an open, close-to-metal software stack~\cite{amdxDNAgithub,hunhoff2025efficiencyexpressivityextensibilityclosetometal,amd_aie_api_ug1529_2025_2}. Built around programmable AI Engine tiles with explicit control over computation and data movement, XDNA2 allows developers to design kernels around the quantization formats that modern LLM users already rely on, rather than forcing the model to conform to a proprietary deployment path.

In this work, we present TileFuse, a close-to-metal mixed-precision kernel library for AMD XDNA2 NPUs that targets GEMM/GEMV-based operators in quantized LLM inference. TileFuse fuses unpacking, dequantization, and execution into a single kernel flow, couples these kernels with weight-layout and dataflow co-design to support practical LLM shapes efficiently, and integrates them into an end-to-end LLM runtime. Our results show that this approach makes XDNA2 a strong accelerator for LLM inference, delivering substantial kernel-level gains and up to $2.0\times$ lower prefilling latency with more than $64.6\%$ lower energy. Our specific contributions are as follows:

\textbf{Fused Kernels for Practical Quantization:}
We design and implement fused AIE kernels that directly consume AWQ-style $W4A16$ and $W8A16$ weights for GEMM/GEMV operations. In this paper, \emph{fusion} means that low-bit weight conversion is integrated directly into the GEMM/GEMV kernel: weights remain in compressed INT4/INT8 format until they reach the compute core, where unpacking, dequantization, and matrix computation are performed within the same kernel flow. By avoiding separate BF16 weight materialization, TileFuse improves kernel-level performance by up to 121.6\% for GEMM and 281\% for GEMV over full-precision baselines.

\textbf{Interleaved Pre-Tiling:} Runtime DMA streaming constraints limit supported weight-matrix dimensions to below 8K, even though the MLP layers of large models often exceed this range. We address this with an interleaved column-major pre-tiling layout that enables efficient runtime streaming and supports GEMM dimensions up to 32K.

\textbf{GEMV Dataflow Co-Design:} 
To accelerate the GEMV-heavy token-generation phase, we introduce a dataflow that distributes weights through memory cores across the full two-dimensional array, improving token-generation throughput.

\textbf{End-to-End Edge Efficiency:} Across Ryzen AI SoCs, our solution outperforms a tuned integrated GPU baseline, achieving up to 2.0$\times$ lower prefilling latency while reducing energy consumption by more than 64.6\%. These results establish XDNA2-based NPUs as a strong accelerator for long-context edge LLM inference.

\textbf{Reconfigurability-Aware NPU Optimization:}
We characterize XDNA2 reconfigurability as a double-edged sword for LLM inference. On one hand, it enables TileFuse to customize compute-core microkernels, DMA-friendly weight layouts, and array-level data movement for different GEMM/GEMV workloads, which is essential for supporting off-the-shelf quantized formats efficiently. On the other hand, runtime dispatch and fabric reconfiguration introduce overhead that must be amortized by sufficiently large operators. TileFuse leverages the advantages of reconfigurability for large prefilling GEMMs and optimized GEMV mappings, while exposing its limitations for latency-sensitive token generation.

The rest of the paper is organized as follows: Section~2 reviews background on LLM inference, quantization, and AMD XDNA2 NPUs. Section~3 discusses several closely related works, their significance, and limitations. Section~4 presents the TileFuse design, including the key challenges, weight pre-tiling strategy, mixed-precision microkernels, GEMV dataflow, and analysis on reconfigurability. Section~5 evaluates TileFuse at the kernel and end-to-end LLM levels. Section~6 discusses generalizability and the implications of hybrid NPU+iGPU deployment. Section~7 concludes.

\section{Background}
\subsection{LLM inference}
LLM inference proceeds in two phases: \emph{prefilling} and \emph{token generation}, which impose different computational workloads and hardware bottlenecks. In prefilling, the full input sequence is processed in parallel, so execution is dominated by matrix multiplication, or GEMM. $C = A \times B$, with $A \in \mathbb{R}^{M \times K}$, $B \in \mathbb{R}^{K \times N}$, and $C \in \mathbb{R}^{M \times N}$. Here, $M$ varies with sequence length, while $N$ and $K$ span a wide range across layers in edge LLMs. We measure prefilling performance using time to first token (TTFT).

In token generation, the model produces one token at a time while reusing the key/value (KV) cache from previous tokens. Because edge inference commonly runs at batch size 1 ($B=1$), many linear layers reduce to GEMV. We measure generation performance using time per output token (TPOT), or equivalently its inverse, tokens per second (tok/s).

\subsection{Quantization}

To reduce model size and relieve memory bottlenecks, on-device LLMs commonly rely on quantization, typically denoted as \emph{w}X\emph{a}Y, where X is the weight precision and Y is the activation precision. In this paper, we consider two widely used formats, $W4A16$ and $W8A16$, with primary emphasis on $W4A16$ because of its popularity in local LLM deployment.

Quantization maps high-precision values to low-bit integers using scale factors and, depending on the quantization scheme, zero-points. A key distinction is between symmetric and asymmetric quantization. In symmetric quantization, real values are mapped around zero, and the integer zero directly represents the real value zero. Therefore, dequantization only requires a scale factor:
\begin{equation}
\hat{w} = s_g q,
\label{eq:symmetric-dequant}
\end{equation}
where $\hat{w}$ is the dequantized weight, $q$ is the quantized value, and $s_g$ is the scale for group or channel $g$.

In asymmetric quantization, the quantized integer range is shifted by a zero-point, allowing the representable range to better match non-zero-centered weight distributions. Dequantization is given by
\begin{equation}
\hat{w} = s_g \bigl( q - z_g \bigr),
\label{eq:asymmetric-dequant}
\end{equation}
where $z_g$ is the zero-point. Asymmetric quantization can improve accuracy for low-bit formats, but it also introduces additional metadata and arithmetic during dequantization.

In the $W4A16$ weight-only format used by AWQ~\cite{lin2023awq}, activations are represented in FP16/BF16, while model weights are stored as packed 4-bit integers with per-group scales and zero-points. Accumulation is typically performed in FP32 to preserve accuracy. Weights are quantized group-wise, so all weights in a group share the same scale and zero-point, which are stored alongside the weights. In our setup, we use AWQ-style $W4A16$ per-group asymmetric quantization with group size 128.

The $W8A16$ format is similar: activations remain in FP16/BF16, weights are stored in INT8, and accumulation is typically performed in FP32. Because INT8 offers higher precision than INT4, $W8A16$ is often deployed in symmetric quantization form. Its quantization parameters are typically applied per channel, so each output channel of the weight matrix shares the same metadata. In our setup, we use BF16 activations with INT8 weights and per-channel symmetric quantization without zero-points.

\vspace{-5pt}
\subsection{Neural Processing Units (NPUs)}

To support local LLM inference under tight power and thermal budgets, edge SoCs increasingly integrate neural processing units (NPUs) alongside CPUs and iGPUs~\cite{amdnpu,intelnpu,qualcommnpu,applenpu}. In principle, NPUs are intended to provide competitive inference performance at higher energy efficiency, but in practice their usefulness for modern LLMs is shaped less by raw compute alone than by their programming model. Across current vendors, NPUs are commonly exposed through proprietary or compiler-driven deployment stacks, such as Apple Core ML~\cite{apple_coreml_2026}, Qualcomm QNN~\cite{qualcomm_qnn_sdk_2025}, Intel OpenVINO~\cite{intel_openvino_npu_2025}, and vendor-managed runtime flows, which simplify deployment but expose limited low-level control over kernels, data movement, and execution policies. As a result, many of these stacks do not natively support practical open-source quantization formats such as AWQ~\cite{lin2023awq}. Instead, they often support only restricted subsets of models and quantization schemes, such as per-tensor, per-channel, or symmetric quantization~\cite{scaling_llm_mobile_2025}, making it difficult to efficiently deploy the asymmetric per-group formats common in modern edge LLMs. Therefore, although NPUs are increasingly present in edge devices, many remain difficult to use as first-class LLM accelerators.

\subsection{AMD XDNA2 and IRON}

AMD integrates its NPU~\cite{amdnpu, rico2024xdna} into Ryzen AI SoCs for laptops, including Strix Point~\cite{amd-ryzen-ai-9-hx-370} and Krackan Point~\cite{amd-ryzen-ai-7-350}. XDNA2 is organized as a tiled spatial array with explicit dataflow and local memory. A typical XDNA2-class NPU consists of 32 AIE compute cores arranged in a $4\times8$ layout, along with 8 memory cores for temporary storage and 8 shim cores that use buffer descriptors (BDs) for efficient data movement (Figure~\ref{fig:AMD_XDNA}).

Unlike the more closed software stacks above, AMD also provides MLIR-AIE (IRON)~\cite{hunhoff2025efficiencyexpressivityextensibilityclosetometal}, an open-source, close-to-metal programming flow that exposes both microkernel design within each compute tile and explicit data movement across the array. This combination of a flexible spatial architecture and an open low-level software stack makes XDNA2 a particularly strong platform for kernel and dataflow research.

\section{Related Works}

\textbf{Quantized LLM inference on CPUs and GPUs.}
As LLMs grow in size, many quantization algorithms have been proposed to compress model weights to low-bit formats while preserving accuracy, including AWQ, GPTQ, and LLM.int8~\cite{lin2023awq,frantar2023gptq,dettmers2022llmint8}. To deploy these quantized models efficiently, GPU-centered open-source edge inference frameworks such as \texttt{llama.cpp}~\cite{gerganov_llamacpp} have become widely used for running LLMs on personal devices. In particular, \texttt{llama.cpp} supports K-Quants~\cite{kawrakow2023kquants}, a family of block-wise quantization formats that store low-bit weights together with block-level scaling metadata to improve the accuracy-efficiency tradeoff of local LLM inference.

Recent systems have further improved quantized LLM inference on GPUs. QServe~\cite{qserve2024} introduces progressive $W4A8KV4$ quantization and system co-design to improve GPU serving efficiency, where $W4$, $A8$, and $KV4$ denote 4-bit weights, 8-bit activations, and 4-bit key/value cache quantization, respectively. MARLIN~\cite{marlin2024} implements high-performance mixed-precision GPU kernels for batched inference. PowerInfer~\cite{powerinfer2024} targets local inference on consumer-grade GPUs by exploiting activation locality to reduce memory pressure and enable efficient GPU-CPU hybrid execution. These works demonstrate the effectiveness of quantization-aware system and kernel design, but they primarily target CPUs and GPUs, where optimized quantized kernels and mature software stacks already exist, rather than client NPUs.

\textbf{LLM inference on mobile and client NPUs.}
As NPUs become common in edge devices, recent works have explored their role in on-device LLM inference. Some systems exploit SoC heterogeneity by studying NPU offloading and scheduling strategies for end-to-end performance~\cite{fast_ondevice_npu_2025,mobile_soc_characterization,powerinfer2_2024}. These works demonstrate the value of NPUs in heterogeneous edge inference, but their main focus is global execution orchestration rather than close-to-metal low-bit NPU kernel design. Another line of work focuses more directly on quantized LLM execution on mobile NPUs. Hao et al.~\cite{scaling_llm_mobile_2025} show that Qualcomm Snapdragon NPUs can support LLM test-time scaling by exploiting underutilized matrix units during decoding. Their system introduces a hardware-aware tile quantization scheme and LUT-based implementations of Softmax and dequantization, enabling parallel test-time scaling on smartphone NPUs. While this is a compelling demonstration of mobile NPU potential, it remains tightly coupled to Qualcomm's platform and to a hardware-shaped quantization/data-layout strategy. In contrast, TileFuse targets AMD XDNA2 NPUs through a close-to-metal programming flow and directly supports practical off-the-shelf AWQ-style $W4A16$ and $W8A16$ formats through reusable fused GEMM/GEMV kernels, rather than requiring a platform-specific quantization scheme.

\textbf{Close-to-metal optimization on AMD XDNA2 and AIE.}
AMD XDNA2-based NPUs, together with the flexible IRON programming flow~\cite{rico2024xdna,hunhoff2025efficiencyexpressivityextensibilityclosetometal}, provide a more open platform for close-to-metal NPU research. Prior work has explored both compiler-level and kernel-level optimizations for AIE-style architectures. AIE4ML~\cite{danopoulos2025aie4ml} and MLIR-AIR~\cite{wang2025mlir_air_amd_npu} focus on compiler and framework support for AIE, enabling AI acceleration through automation and graph-level mapping optimizations. Other works study low-level optimization and end-to-end applications on AIE/XDNA architectures. MaxEVA~\cite{taka2023maxeva} applies several low-level matrix-multiplication optimizations on Versal AIE, demonstrating the performance and energy-efficiency potential of AIE devices. More recently, Taka et al.~\cite{taka2025striking_balance} and Wang et al.~\cite{wang2025asymmetrictilebufferingbeneficial} study generic GEMM optimization on Ryzen AI NPUs and XDNA2, introducing systematic optimization methodology and asymmetric tile buffering to improve throughput. In particular, Wang et al. show that asymmetric tile buffering can reach BFP16--BF16~\cite{song2018computation} GEMM performance up to 28.5 TFLOPS on XDNA2, where BFP16 denotes a block-floating-point datatype in which a block of values shares one exponent. However, these works focus on high-throughput generic GEMM or BFP-style formats, rather than practical low-bit LLM quantization schemes such as AWQ-style $W4A16$, and they do not target end-to-end deployment of off-the-shelf quantized LLMs.

FastFlowLM~\cite{fastflowlm_github_2026} is a recent runtime system for deploying LLMs on AMD Ryzen AI NPUs and demonstrates the practical potential of using XDNA-class NPUs for local LLM inference. However, FastFlowLM is primarily presented as a deployment/runtime framework, with limited public detail on its low-level kernel implementation and optimization methodology. TileFuse is complementary but targets a different goal: it provides an open, reusable fused mixed-precision kernel library and explains the underlying kernel/dataflow design, including in-core dequantization, metadata-aware offline pre-tiling, large-matrix streaming support, and topology-aware GEMV dataflow. Therefore, TileFuse exposes the close-to-metal mechanisms needed to support standard quantized LLM formats on XDNA2, rather than only providing a black-box runtime interface.

The closest prior AMD-NPU application work is \textit{Mapping Gemma3 onto an Edge Dataflow Architecture}~\cite{du2026gemma3_edge_dataflow}, which presents end-to-end deployment of quantized Gemma3 models on Ryzen AI NPUs with hardware-aware dequantization, tiled matrix multiplication, and fused decoding kernels. In contrast, TileFuse focuses on a reusable library of fused mixed-precision GEMM/GEMV kernels for practical quantized LLM inference on XDNA2. TileFuse directly targets off-the-shelf AWQ-style $W4A16$ and $W8A16$ formats, bridging the gap between XDNA2's raw kernel potential and practical deployment of standard quantized LLMs on client NPUs.

\section{TileFuse Overview}

TileFuse provides an end-to-end, close-to-metal implementation that can be compiled using the MLIR-AIE (IRON) flow and directly executed on AMD XDNA2 NPUs. We focus on GEMM- and GEMV-based operators in LLM inference, which dominate prefilling and token generation, respectively (batch size = 1). Our design has three components. First, weight pre-tiling packs quantization metadata with weights and reorganizes weight layout for efficient streaming and large-layer support. Second, mixed-precision microkernels fuse unpacking, dequantization, and GEMM/GEMV while preserving weight reuse. Third, a GEMV-specific dataflow redesign improves utilization across the full 4$\times$8 AIE array during generation-side linear layers.

\subsection{Design Challenges}
Although AMD's XDNA2 NPU provides a programmable spatial architecture, efficient execution of quantized LLM workloads on this platform remains difficult. The challenge is not only to run low-bit kernels, but to do so in a way that matches practical LLM quantization formats and co-optimizes data movement and computation. In our target setting, these difficulties arise most prominently in transformer linear layers, which dominate the prefilling phase through large GEMMs and the token generation phase through GEMV-like operations. We identify four main design challenges.

\subsubsection{Efficient On-the-Fly Dequantization.}
A straightforward mixed-precision execution flow separates dequantization from matrix multiplication: quantized weights are first read from memory, expanded into BF16, written back to memory, and then reloaded by a GEMM or GEMV kernel. This avoids modifying the existing full-precision GEMM/GEMV kernels, but introduces an unnecessary global-memory round trip for the intermediate dequantized weights.

Fusing dequantization into GEMM/GEMV removes this global-memory materialization, but it also introduces a new reuse challenge. In tiled GEMM, the same weight tile is typically multiplied with multiple activation tiles along the rows of the activations. Therefore, if the fused kernel simply dequantizes the weight tile every time it is consumed by an inner matrix-multiply operation, the global-memory traffic is reduced, but the same dequantization work is repeated many times. This is inefficient because the dequantized weight tile should be reused after it is produced.

Thus, the key challenge is not only to perform dequantization on the fly, but also to preserve the reuse behavior of tiled GEMM/GEMV. On XDNA2, this requires explicitly managing the limited compute-core local memory: the kernel should dequantize each quantized weight tile once, store the BF16 result in local memory, and reuse it for subsequent matrix-multiply operations before moving to the next weight tile.

\subsubsection{Feeding Metadata Under Limited Streams}
Beyond the fused arithmetic itself, a mixed-precision kernel must also deliver all the required data into the compute tiles efficiently. To dequantize weights on the fly inside compute cores, the compute tile must access the weights' corresponding metadata (scale and, for asymmetric quantization, zero-point). 

However, the XDNA2 architecture provides limited streaming interfaces to each compute tile~\cite{amd_ug1079_interface_considerations_2025_2}, making it impractical to supply weights, activations, and metadata as three fully independent runtime streams. As a result, supporting practical quantization formats is not merely a matter of arithmetic conversion inside the microkernel; it also requires a data layout that makes the required metadata available at the right time and in the right place without introducing additional streaming overhead.

\subsubsection{Supporting Large MLP Layers}
LLM MLP layers often involve very large weight matrices, with dimensions easily reaching or exceeding the scale of typical tiled kernels. In the default tiled layout, compute cores are assigned weight tiles in an interleaved column pattern across the AIE array. After processing tiles from output column $a$, the next required tiles reside in column $a+8$, implying a large stride in memory. 
Because the DMA stride register has limited capacity, when the matrix is large, this stride can be too large to be streamed through a single buffer descriptor (BD). It forces a single GEMM operation to be split across multiple kernel invocations, incurring significant extra overhead. Therefore, supporting realistic LLM layer shapes requires more than functional correctness: it requires a weight organization strategy that preserves efficient streaming across a wide range of matrix sizes.

\subsubsection{Hardware Underutilization for GEMV}

While prefilling is dominated by large GEMMs, token generation in edge LLM inference frequently operates at batch size 1, making many linear layers GEMV-like. Applying a mixed-precision kernel to GEMV introduces a different optimization challenge: compared with full-precision GEMV, transmitting quantized weights leads to less bandwidth demand, while on-the-fly dequantization increases compute work. However, the existing GEMV kernel uses a straightforward broadcast strategy that does not fully exploit the available 4$\times$8 compute array, significantly underutilizing the hardware. This combination shifts GEMV from being primarily memory-bound to being more compute-bound, making computation throughput critical. A GEMV implementation must address both microkernel efficiency and array-level data distribution. Without such co-design to improve computation throughput, the token-generation kernel remains inefficient even if quantized GEMM execution is improved.

These challenges motivate a co-designed solution across weight layout, metadata placement, microkernel structure, and array-level dataflow, which we present next.

\subsection{Weight Pre-tiling}
\begin{figure}[t]
  \centering
  \includegraphics[width=1\linewidth]{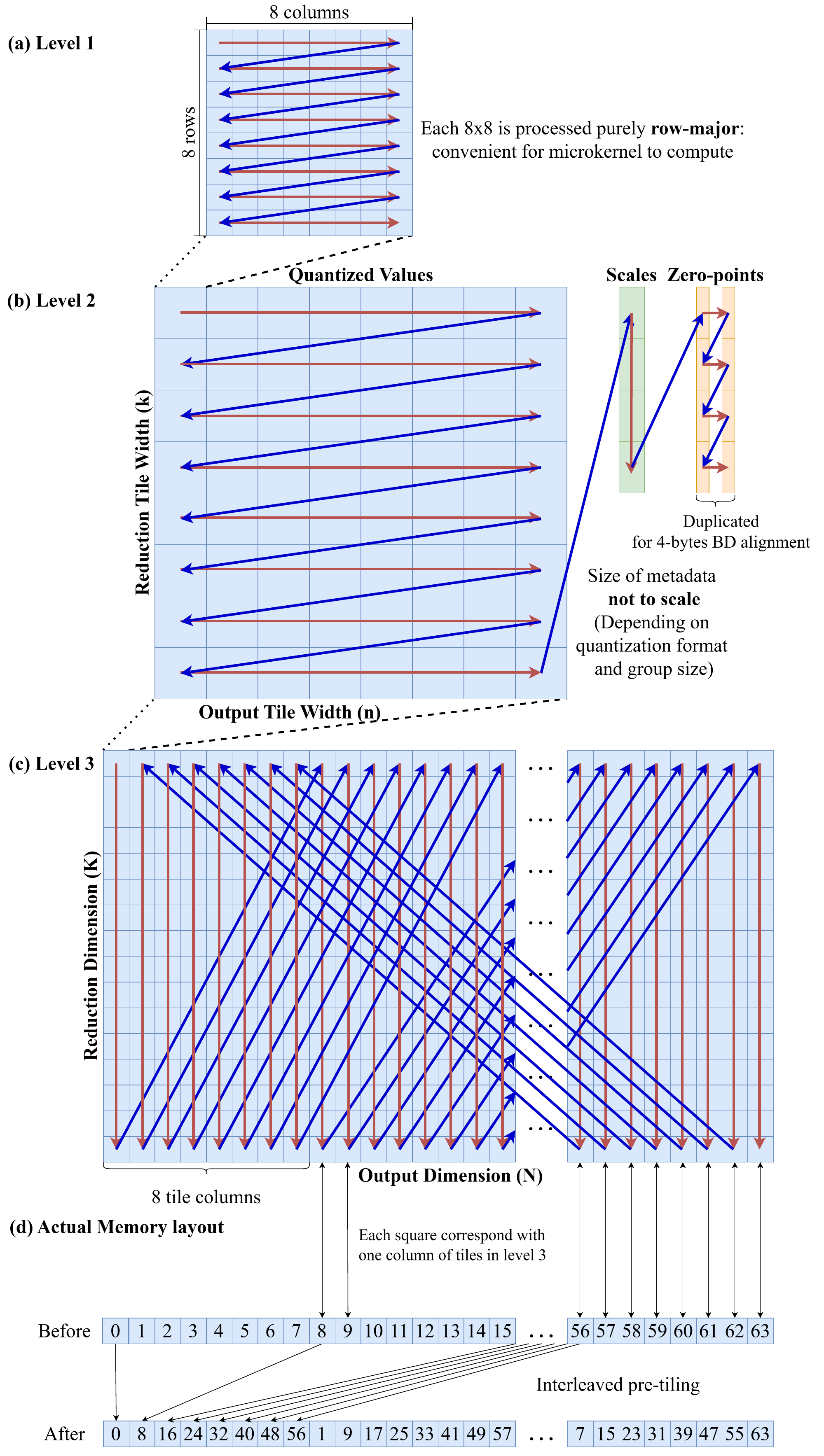}
  \caption{The pre-tiled layout of the weight matrix for $W4A16$ quantization format on three levels. (a) The row-major layout of each 8$\times$8 weight block. (b) The pre-tiling and metadata packing for each $k\times n$ tile. (c) The interleaved column-major pre-tiling layout on the full input matrix. (d) Actual memory layout before and after interleaved pre-tiling.}
  \label{fig:Pre-tiling}
  \vspace{-5pt}
\end{figure}

\begin{figure*}[t]
  \centering
  \includegraphics[width=0.8\linewidth]{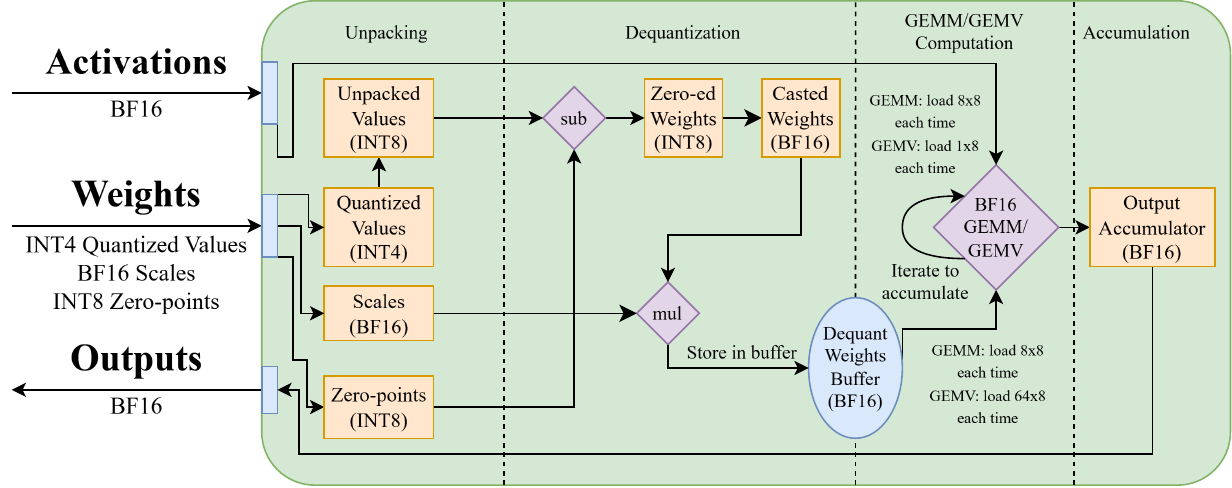}
  \vspace{-5pt}
  \caption{The mixed-precision GEMM/GEMV microkernel design for $W4A16$ quantized weight, fusing unpacking, dequantization, GEMM/GEMV computation, and accumulation.}
  \label{fig:Microkernel}
  \vspace{-5pt}
\end{figure*}

Because model weights are known at deployment time, we perform offline pre-tiling on quantized weights with no runtime overhead. This pre-tiling step serves two purposes: it packs each weight tile with the metadata required for dequantization, and it reorganizes the weight layout to improve streaming efficiency under DMA BD constraints. As a result, pre-tiling enables both metadata-aware mixed-precision execution and efficient support for large MLP layers. The pre-tiling layout is illustrated in Figure~\ref{fig:Pre-tiling} with three levels: Level 1 shows the layout within each $8\times8$ weight block, 
Level 2 shows how blocks and metadata are packed within each $k\times n$ tile, 
and Level 3 shows how tiles are arranged across the full $K\times N$ weight matrix. Arrows indicate memory continuity between tiles/data, with the purple arrow representing continuity within the same row/column and blue arrows representing continuity when switching row/column.

\textbf{Packing weights with quantization metadata.} 
In a regular GEMM kernel, the input matrices are divided into small tiles for each compute core to process. Therefore, we pre-tile the weight data and pack metadata directly into each pre-tiled weight block in memory. 
 
For $W4A16$, as shown in Figure~\ref{fig:Pre-tiling}b, each quantized weight tile is stored in a way such that all packed INT4 values of the tile are stored contiguously first, followed by the BF16 scales and INT8 zero-points for the quantization groups in that tile. This enables the compute core to fetch all required information from the weight stream. For example, a $128\times64$ (128 rows and 64 columns) weight tile needs 64 scales and 64 zero-points for dequantization because with a group size of 128, every 128 weights in the same column share the same scale and zero-point. Therefore, we pack them together such that the $128\times64$ INT4 weights are followed by $64$ BF16 scales, which are again followed by $128$ INT8 zero-points (duplicated 64 zero-points)\footnote{Because the buffer descriptors (BD) transfer data through DMA engines, we duplicate the zero-point array so that each tile payload size is divisible by 128 bytes, as required by the DMA hardware. The additional memory footprint is negligible, taking only 1.5\% of extra memory.}. In $W8A16$, symmetric INT8 weights only require scales to dequantize, which are appended in the same manner during pre-tiling. 
While some metadata may be replicated during pre-tiling, the extra metadata is negligible compared to the weight traffic saved by INT4/INT8 streaming.

\textbf{Interleaved pre-tiling for large-matrix support.}
For large LLM layers, especially MLP projections, the weight matrix can exceed the stride limits supported by the runtime BD programming model. Because the AIE array has 8 compute-core columns, the output-tile columns are distributed across these 8 columns in a round-robin manner. For example, AIE column 0 processes output-tile columns $0, 8, 16, \ldots$, AIE column 1 processes output-tile columns $1, 9, 17, \ldots$, and so on. Therefore, for a given AIE column, consecutive tiles in its execution order are separated by 8 tile columns in the original row-major weight layout. This creates a large memory stride when the weight matrix is stored in the default layout. To eliminate this large-stride access, we introduce an interleaved column-major pre-tiling layout that places tiles assigned to the same AIE column consecutively in memory, as illustrated in Figures~\ref{fig:Pre-tiling}c and~\ref{fig:Pre-tiling}d. As a result, the BD can access data contiguously without large strides in memory, enabling the kernel to support matrices with dimensions $K$ and $N$ up to 32k without losing performance.

\textbf{Tiled layout for bandwidth-efficient streaming.} 
Beyond metadata packing and large-matrix support, pre-tiling alleviates the memory bandwidth bottleneck by converting the original row-major weight matrix into a tile-major layout, matching how the hardware kernel consumes the input. This enables DMA engines to issue long contiguous reads and stream weights into the NPU, which better utilizes DRAM burst transfers and reduces wasted bandwidth. Avoiding tiling during runtime also helps maintain consistent performance across different GEMM shapes, as shown in Section~5.

\vspace{-10pt}

\subsection{Mixed-precision Microkernels}

Given the pre-tiled weight layout, we implement mixed-precision microkernels that fuse unpacking, dequantization, and GEMM/GEMV inside each compute core. The goal is to perform on-the-fly dequantization using registers and local memory while preserving throughput.

\textbf{GEMM microkernel.} Figure~\ref{fig:Microkernel} illustrates the fused mixed-precision microkernel flow. AIE matrix-multiply primitives are optimized for BF16 $8\times8$ blocks, so we divide the $m\times k$ activation tile and the $k\times n$ weight tile into $8\times8$ sub-tiles (Figure~\ref{fig:Pre-tiling}a) and accumulate partial results into accumulators. For $W4A16$, the input weight stream contains packed INT4 values, with two 4-bit weights stored in each byte. As shown in Figure~\ref{fig:Microkernel}, the microkernel first performs \emph{unpacking}: it extracts the low and high 4-bit fields using bit masking and shifting, and widens them into INT8 lanes so that each quantized weight can be processed as an individual integer value. This unpacking step does not yet recover the numerical BF16 weight; it only converts the packed storage format into an integer vector format suitable for arithmetic.

After unpacking, the widened INT8-represented quantized values are converted to BF16 and dequantized using the corresponding scale and zero-point metadata following Eq.~(\ref{eq:asymmetric-dequant}). For symmetric $W8A16$, the INT8 weight value is directly converted to BF16 and multiplied by the per-channel BF16 scale, without zero-point subtraction, following Eq.~(\ref{eq:symmetric-dequant}). The resulting BF16 weights are stored in a compute-core local buffer and then consumed by the AIE matrix-multiply primitives for GEMM/GEMV operations.

A naive fusion approach would be to apply unpacking and dequantization every time an $8\times8$ weight block is loaded from the input. However, the same weight block is reused across multiple activation slices, so dequantizing on every load is redundant and significantly reduces microkernel throughput. To avoid redundant dequantization, we leverage the local memory within each compute core: the kernel unpacks and dequantizes each quantized weight block once, stores the resulting BF16 weights into the local buffer, and reuses them for subsequent GEMM computation. Although this method introduces extra load and store operations because of the use of a local memory buffer, the VLIW instructions can hide most of the memory overhead by overlapping memory operations with dequantization and GEMM computation~\cite{amd_ug1079_aie_arch_overview_2025_2}.

\textbf{GEMV microkernel.} We apply the same fusion strategy to GEMV and implement mixed-precision GEMV microkernels that directly consume quantized weights. However, GEMV requires additional optimization to improve compute utilization. Compared with full-precision GEMV, transmitting quantized weights leads to less bandwidth demand, while on-the-fly dequantization increases compute work. This combination shifts GEMV from being primarily memory-bound to being more compute-bound, making arithmetic throughput within the microkernel critical.

The baseline GEMV microkernel does not fully utilize the compute core’s arithmetic resources: it repeatedly loads $16\times8$ weight blocks (two 8x8 weight blocks), and uses vector operations with activation fragments to accumulate only $16$ output elements per iteration, which underutilizes the vector hardware for GEMV. Since a vector register can hold up to $64$ BF16 values~\cite{amd_aie_api_ug1529_2025_2}, we load a $64\times8$ weight block (eight 8x8 weight blocks) into these registers per iteration from local buffer to perform GEMV with the $1\times8$ activations, accumulating $64$ output elements per iteration. The $1\times8$ activation fragment corresponds to one slice of the input vector along the $K$ dimension. This same activation fragment is reused across $64$ different weight rows, each of which corresponds to a different output element. Therefore, multiplying a $64\times8$ weight block with a shared $1\times8$ activation fragment produces $64$ partial output accumulations. As the microkernel iterates over the $K$ dimension, these partial sums are accumulated into a $64\times1$ output vector. By widening the inner microkernel to increase work per instruction and reuse each activation fragment across more output rows, we improve per-core utilization and throughput for token generation workloads.

\vspace{-5pt}

\subsection{Dataflow for Weight Distribution}

\begin{figure}[t]
\centering
\includegraphics[width=1\linewidth]{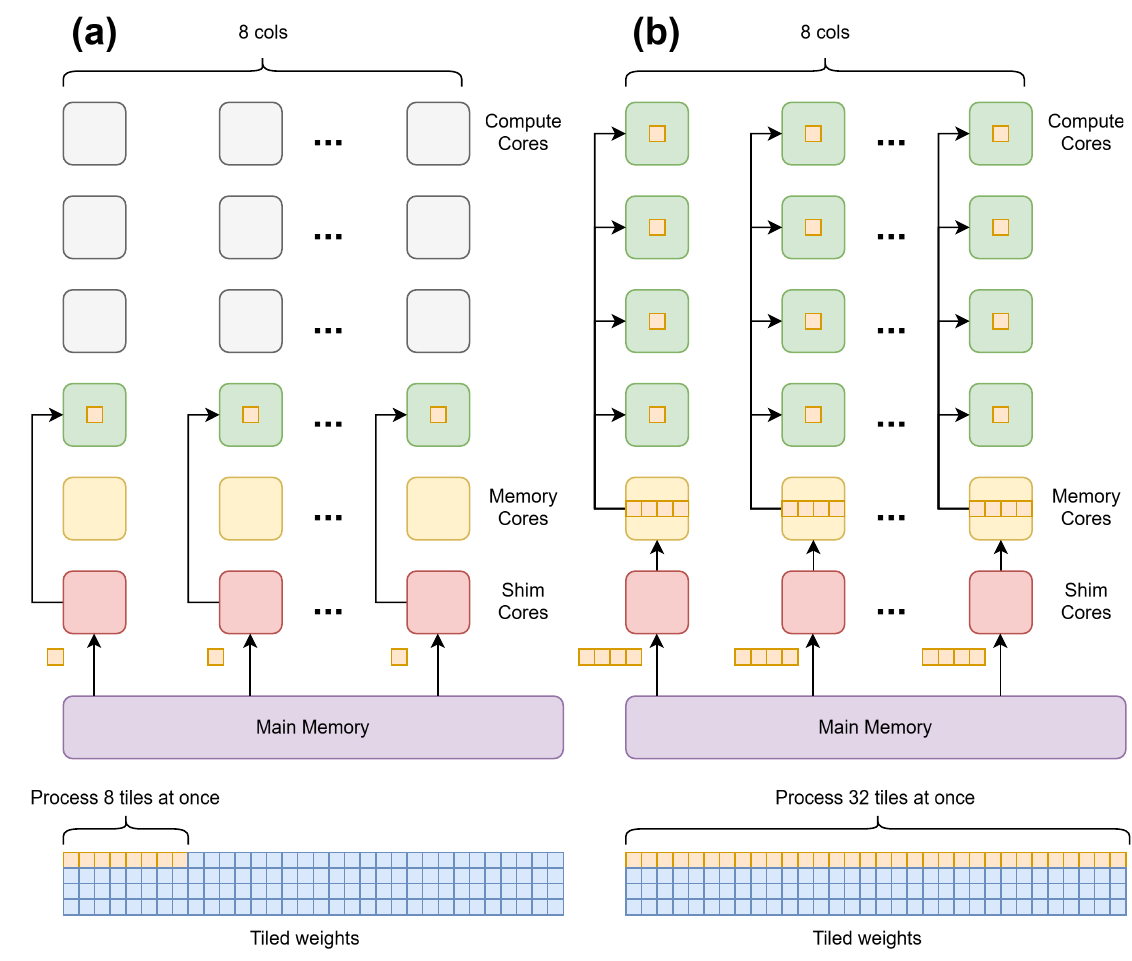}
\caption{Comparison of (a) baseline GEMV dataflow utilizing a single compute row and (b) optimized GEMV dataflow distributing weights to all cores to maximize utilization.}
\label{fig:Dataflow}
\end{figure}

As discussed in Section~4.3, switching from full-precision to mixed-precision reduces weight bandwidth demand but adds unpacking and dequantization work, making GEMV more compute-bound rather than memory-bound. However, the baseline data movement pattern still feeds only one row of compute cores, leaving most of the $4\times8$ AIE array idle. We therefore redesign the GEMV dataflow, optimizing the spatial movement of input data through shim cores, memory cores, and compute cores.

In the baseline GEMV mapping shown in Figure~\ref{fig:Dataflow}a, each shim column streams weights directly to the compute cores in one compute row. This direct streaming path is simple, but it only uses $8$ of the $32$ compute cores. Extending this direct path to all four compute rows is difficult because GEMV requires each compute core to consume a distinct weight tile. The same activation vector can be broadcast to all compute cores, but each compute core must receive a different $n\times k$ weight tile to compute distinct output elements. Since each AIE column has only one shim core connected to external memory, the shim core cannot naturally provide four independent weight streams to the four compute rows in the same column.

Our optimized mapping in Figure~\ref{fig:Dataflow}b uses the memory core as an intermediate distribution layer. The weight-delivery dataflow has two stages. In the first stage, each shim core streams a contiguous bundle of four weight tiles from external memory into its associated memory core, where the four tiles correspond to the four compute rows in that AIE column. In the second stage, the memory core distributes these four tiles to the four compute cores in the column. The activation vector is shared across the array, while each compute core receives a distinct weight tile and computes a distinct portion of the output vector.

Importantly, intermediate GEMV accumulation results do not flow through these stages. Each compute core performs unpacking, dequantization, multiplication, and accumulation locally for its assigned output elements. The memory core is used only for input weight distribution, not for reducing partial sums across compute cores. Therefore, by using memory cores as an intermediate stage for weight distribution, this new design increases spatial utilization: instead of processing $8$ output-tile groups at a time using one compute row, the optimized scheme processes $32$ output-tile groups at a time using all four compute rows. Although the memory-core distribution stage adds local data movement, this overhead is offset by the increased compute parallelism and is evaluated in the GEMV ablation study in Section 5.5.

\subsection{Leveraging XDNA2 Reconfigurability}

A central reason TileFuse can support practical quantized LLM workloads is that XDNA2 exposes configurable control over both compute and data movement. In this work, reconfigurability does not mean changing the hardware itself or dynamically changing the routing fabric while a kernel is running. Instead, it refers to configuring the AIE array with workload-specific compute-tile programs, memory layouts, DMA buffer-descriptor access patterns, synchronization objects, and inter-core data-movement paths generated by the IRON/MLIR-AIE flow. TileFuse leverages this flexibility at three levels.

First, at the compute-core level, we replace a fixed full-precision GEMM/GEMV datapath with mixed-precision microkernels that fuse unpacking, dequantization, and matrix computation. This compute-tile programmability allows the core to directly consume compressed INT4/INT8 weights, convert them to BF16 locally, and reuse the dequantized weights through local memory. Without the ability to customize the microkernel, the implementation would have to materialize BF16 weights before invoking a standard GEMM/GEMV kernel.

Second, at the memory-layout and DMA level, we use offline pre-tiling to match the runtime access pattern of the AIE array. The interleaved column-major layout changes the physical order of weight tiles so that tiles assigned to the same AIE column become contiguous in memory. This allows the BD-based DMA path to stream large GEMM weights efficiently without large strides or repeated kernel invocations. In this sense, TileFuse uses XDNA2 configurability not only for arithmetic, but also for adapting memory movement to the shape of LLM weight matrices.

Third, at the array level, we configure a GEMV-specific weight-distribution path through memory cores. The baseline direct streaming pattern uses only one compute row, but token-generation GEMV requires each compute core to consume distinct weight tiles. By routing bundled weight tiles from shim cores through memory cores to multiple compute rows, TileFuse maps GEMV across the full $4\times8$ compute array and improves spatial utilization. This is the main optimization in TileFuse that relies on array-level data-movement configuration.

These benefits come with an important limitation. NPU kernel dispatch and AIE configuration introduce runtime overhead, including the cost of launching a precompiled kernel and activating its compute programs, DMA descriptors, synchronization objects, and data-movement connections. This overhead is amortized well by large prefilling GEMMs but becomes costly for short GEMV-like operators during token generation, as discussed in the evaluation and discussion sections. Therefore, TileFuse uses XDNA2 reconfigurability where it provides clear benefits---custom fused kernels, large-matrix streaming, and full-array GEMV mapping---while the evaluation also shows that this flexibility is less favorable for latency-sensitive small operators.

\vspace{-10pt}

\subsection{End-to-End LLM Setup with NPU Kernels}
To evaluate our kernels in real LLMs for both correctness and performance, we built a custom framework in C++ with Python bindings. The Python frontend loads model weights from Hugging Face \texttt{safetensors}, applies our offline weight pre-tiling and metadata packing, and handles tokenization and output decoding. The libtorch-based C++ backend implements the transformer execution loop and dispatches operators to the appropriate device.

In each transformer block, linear layers such as attention projections and MLP projections are dispatched to the NPU when using the TileFuse path. These linear operators call our precompiled XDNA2 kernels with pre-tiled quantized weights and BF16 activations. Nonlinear and attention-related operators, including FlashAttention-2~\cite{dao2023flashattention2}, softmax, residual addition, normalization, and other lightweight element-wise operations, remain on the iGPU using optimized HIP/libtorch kernels. This division follows the compute characteristics of the devices: large GEMM operators can amortize NPU dispatch and reconfiguration overhead, while latency-sensitive and irregular operators are better handled by the iGPU software stack. We adopt this hybrid design based on the compute characteristics of XDNA2~\cite{rico2024xdna} and our profiling of transformer attention blocks.

Our current implementation does not execute the NPU and iGPU concurrently on different operators. Instead, execution proceeds sequentially according to the transformer data dependencies: the host dispatches an operator to either the NPU or the iGPU, waits for its result, and then launches the next dependent operator. Therefore, the hybrid setup should be interpreted as operator-level device selection rather than parallel NPU--iGPU co-execution. We adopt this design to isolate the benefit of TileFuse for quantized linear layers while relying on mature GPU kernels for the remaining transformer operations.

\begin{figure*}[t!]
  \centering

  \begin{subfigure}[t]{0.49\textwidth}
    \centering
    \includegraphics[width=\linewidth]{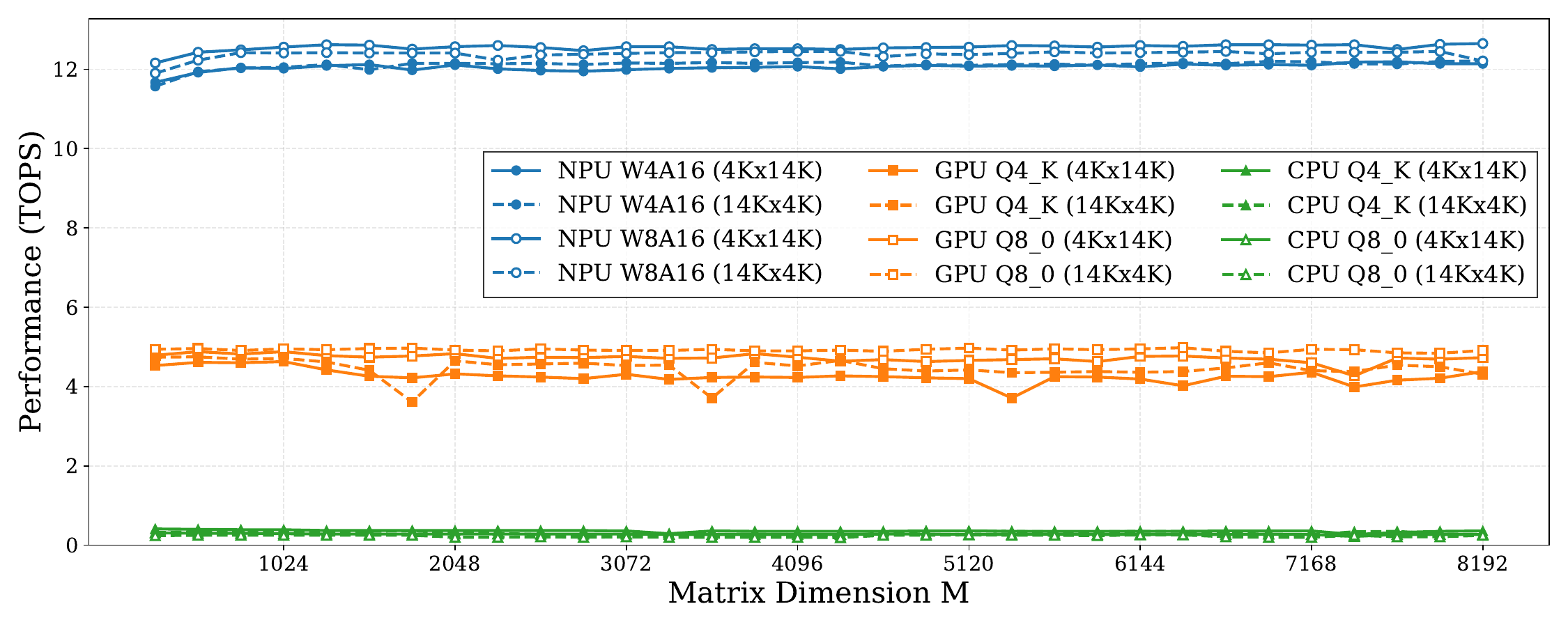}
    \caption{AMD Ryzen AI 7 350: Krackan Point}
    \label{fig:krackan_kernels}
  \end{subfigure}
  \begin{subfigure}[t]{0.49\textwidth}
    \centering
    \includegraphics[width=\linewidth]{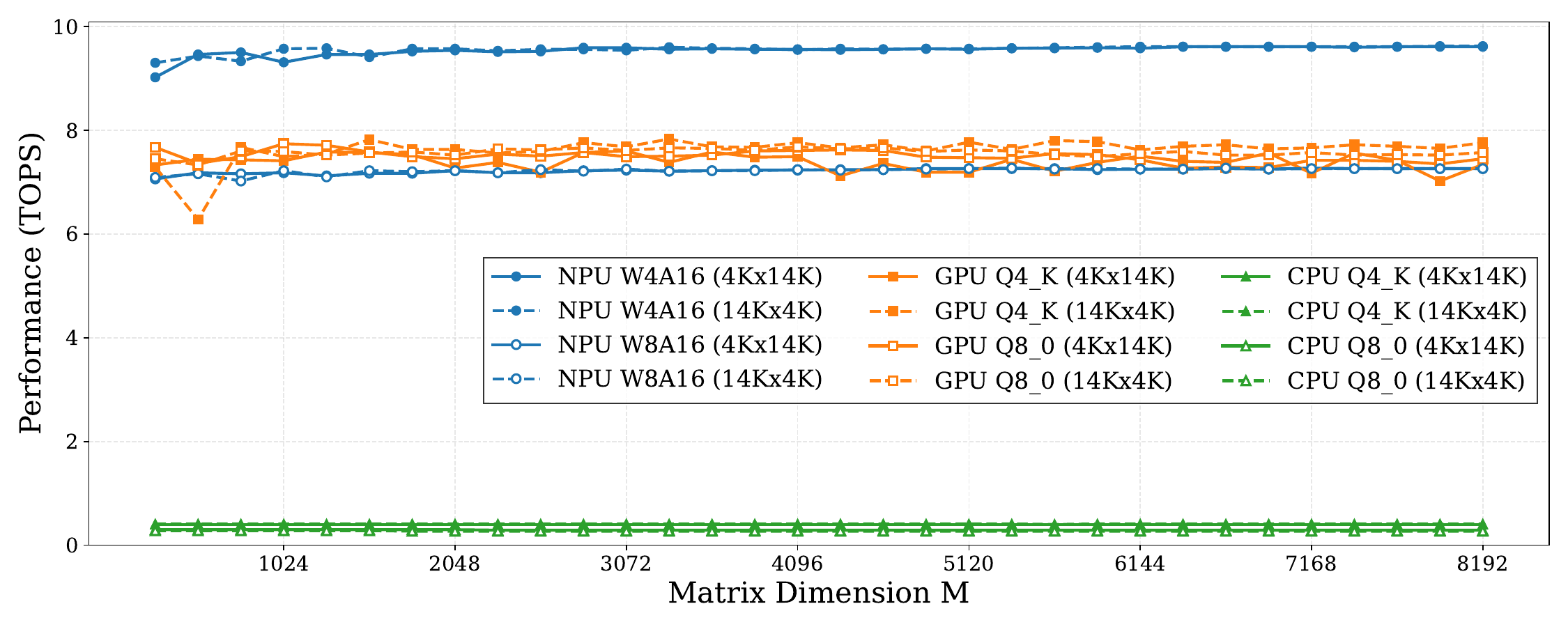}
    \caption{AMD Ryzen AI 9 HX 370: Strix Point}
    \label{fig:strixpoint_kernels}
  \end{subfigure}\hfill
\caption{GEMM throughput across matrix sizes for CPU, iGPU, and NPU quantized kernels. CPU and iGPU curves use \texttt{llama.cpp}'s \texttt{Q4\_K} and \texttt{Q8\_0} fused quantized kernels. NPU curves use TileFuse $W4A16$ and $W8A16$ mixed-precision kernels.}
  \label{fig:kernel_comparison}
\end{figure*}

\section{Evaluation}
\subsection{Kernel Evaluation}

To evaluate kernel performance and shape sensitivity, we profile our $W4A16$ AWQ~\cite{lin2023awq} and $W8A16$ LLM.int8~\cite{dettmers2022llmint8}-style kernels, using group size 128 and per-channel quantization, respectively. Using standard GEMM notation with $A \in \mathbb{R}^{M \times K}$ and $B \in \mathbb{R}^{K \times N}$, we evaluate two configurations with transposed $K$ and $N$ dimensions while sweeping $M$ in steps of 256.

We evaluate TileFuse on two Ryzen AI platforms: Ryzen AI 7 350 and Ryzen AI 9 HX 370. Both platforms include XDNA2-class NPUs and have similar nominal peak memory bandwidth, up to approximately 128~GB/s under comparable high-speed memory configurations~\cite{amd-ryzen-ai-7-350,amd-ryzen-ai-9-hx-370}. However, their iGPU configurations differ substantially: the Ryzen AI 9 HX 370 integrates Radeon 890M graphics with 16 graphics cores, while the Ryzen AI 7 350-class platform uses Radeon 860M graphics with 8 graphics cores~\cite{amd-ryzen-ai-9-hx-370,amd-ryzen-ai-7-350}. Therefore, the HX 370 provides a stronger iGPU baseline primarily because of its larger iGPU, while the AI 7 350 offers a more informative comparison for isolating the benefit of NPU offloading against a smaller integrated GPU.

As a state-of-the-art baseline, we use the fused quantized HIP kernels from \texttt{llama.cpp}~\cite{gerganov_llamacpp} on the iGPU. These were the highest-performing quantized GEMM kernels we found for AMD Ryzen iGPUs as of January 2026, especially after a recent update that improved performance by up to 1.5$\times$ over 2025 releases. We compare against the \texttt{Q4\_K} and \texttt{Q8\_0} formats.\footnote{\texttt{Q4\_K} and \texttt{Q8\_0} are native \texttt{llama.cpp} block-quantized weight formats: \texttt{Q4\_K} stores weights at 4 bits per value using the K-quant layout with additional shared scale/offset metadata, while \texttt{Q8\_0} stores weights at 8 bits per value with block-level scaling. In both cases, \texttt{llama.cpp} fuses dequantization with the HIP matrix kernels. On RDNA iGPUs, these kernels can exploit $dp4a$ instructions~\cite{amd2024rdna35isa}, which are GPU vector dot-product instructions that accumulate multiple low-precision integer products per instruction. This makes the quantized path more compute- and bandwidth-efficient than executing the same operation through unfused floating-point kernels.}

\begin{figure}[t!]
\centering
\includegraphics[width=\linewidth]{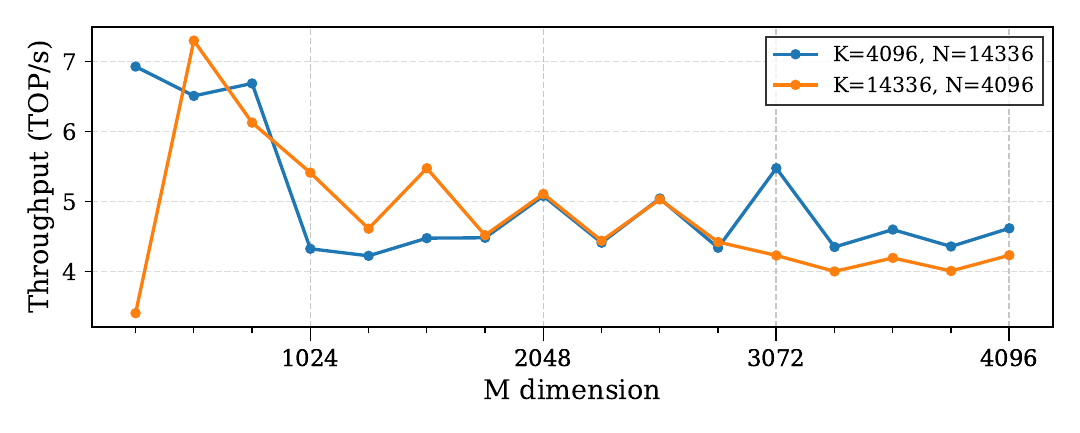}
\caption{Illustrative zig-zag shape sensitivity of intentionally untuned iGPU GEMM kernels on the Ryzen AI 7 350. These kernels are deliberately not optimized for the 8-CU iGPU and are therefore not used as best-performance baselines; instead, the plot shows how suboptimal thread-to-CU mapping can produce non-monotonic throughput across $M$ for two GEMM shapes with swapped $K$ and $N$ dimensions.}
\label{fig:gpu_zigzag}
\end{figure}

Figure~\ref{fig:kernel_comparison} shows that our NPU kernels produce stable performance across shapes on XDNA2. Some accelerators exhibit strong shape sensitivity for GEMM because different matrix dimensions can map unevenly onto the available execution resources. For example, when the number of tiles or thread blocks does not align well with the number of compute units, some resources may remain underutilized, leading to zig-zag throughput as the matrix dimension changes. In contrast to zig-zag patterns such as the untuned iGPU example shown in Figure~\ref{fig:gpu_zigzag} and the more severe TOPS degradation reported for Snapdragon Gen 3~\cite{mobile_soc_characterization},\footnote{In the cited Snapdragon Gen 3 study, GEMMs with orientation $([4096,14336]\times[14336,K])^{T}$ fall to roughly the 1-TOPS range, while the flipped orientation reaches roughly the 6-TOPS range despite comparable arithmetic work.} we observe smooth scaling, which we attribute to our custom weight layout that maintains efficient data streaming under different sizes. Krackan Point and Strix Point reach peak throughput of 12 TOPs and 9 TOPs, respectively, although Strix Point shows lower throughput for the less compact $W8A16$ kernels. Given the shared XDNA2 architecture, this gap likely reflects platform-level factors such as silicon quality, thermal design, or effective NPU-side memory behavior rather than compute scaling alone.

For comparison, iGPU throughput scales roughly with the number of compute units (CUs), the primary parallel execution blocks in AMD RDNA GPUs. The 8-CU Krackan Point iGPU reaches about 5 TOPS, while the 16-CU Strix Point iGPU reaches about 7 TOPS; in contrast, CPU throughput remains marginal at roughly 200 GOPS. We also observe periodic throughput spikes for some GEMM shapes on the iGPU, which likely stem from inefficient thread-to-CU mapping and shape-dependent occupancy effects.

\begin{table}[t]
\centering
\renewcommand{\arraystretch}{1.3} 
\begin{tabular}{lcccc}
\hline
\textbf{Processor} & \textbf{Dtype} & \textbf{NPU {\footnotesize (GOPs)}} & \textbf{CPU {\footnotesize (GOPs)}} & \textbf{iGPU {\footnotesize (GOPs)}} \\ \hline
\multirow{2}{*}{\shortstack[c]{Ryzen AI\\7 350}}    & 4-bit & 206.21 & 298.86 & 215.69 \\
                                                    & 8-bit & 104.16 & 140.14 & 145.45 \\ \hline
\multirow{2}{*}{\shortstack[c]{Ryzen AI\\HX 370}}   & 4-bit & 181.22 & 301.76 & 239.48 \\
                                                    & 8-bit & 96.95 & 141.74 & 133.10 \\ \hline
\end{tabular}
\caption{GEMV performance ($1 \times 4096 \times 14336$) measured in GOPs across Ryzen AI platforms comparing 4-bit and 8-bit precisions.}
\label{gemv_comparison}
\end{table}

Finally, results for General Matrix-Vector Multiplication (GEMV) (Table \ref{gemv_comparison}) show consistent performance across various CPU and iGPU configurations, with the CPU being more performant and close to the results derived by the theoretical DRAM bandwidth limit. In contrast, the NPU is optimized for compute rather than memory bandwidth, so its GEMV performance is lower; Krackan Point only slightly outperforms Strix Point on these tasks for the NPU.

\subsection{End-to-End LLM Performance}





\begin{figure}[t!]
  \centering

  \newcommand{\trimplot}[1]{%
    \scalebox{1}[0.80]{%
      \raisebox{-0.5\height}{%
        \includegraphics[width=\linewidth,trim=8 8 4 8,clip]{#1}
      }%
    }%
  }

  \begin{subfigure}[b]{\linewidth}
    \centering
    \trimplot{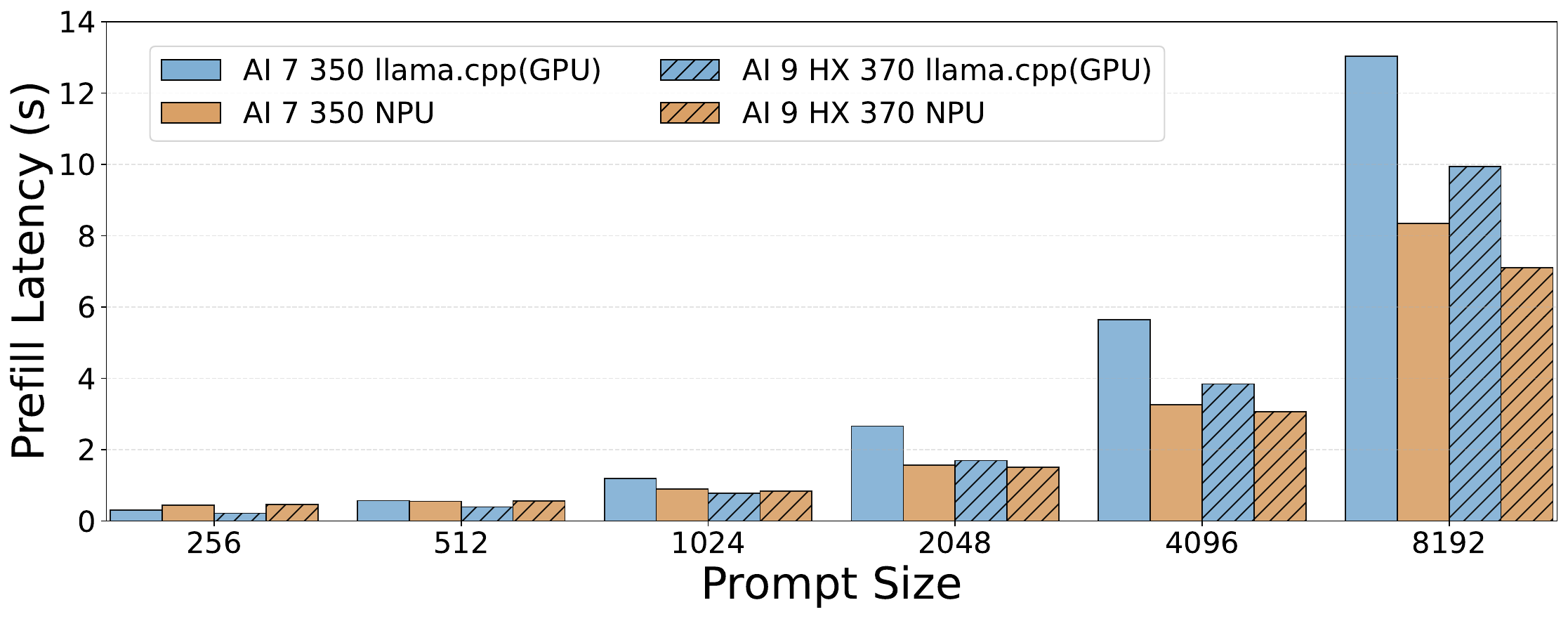}
    \vspace{-5pt}
    \caption{Gemma-2B}
    \label{fig:e2eGemmaprefil}
  \end{subfigure}

  \vspace{0.5em}

  \begin{subfigure}[b]{\linewidth}
    \centering
    \trimplot{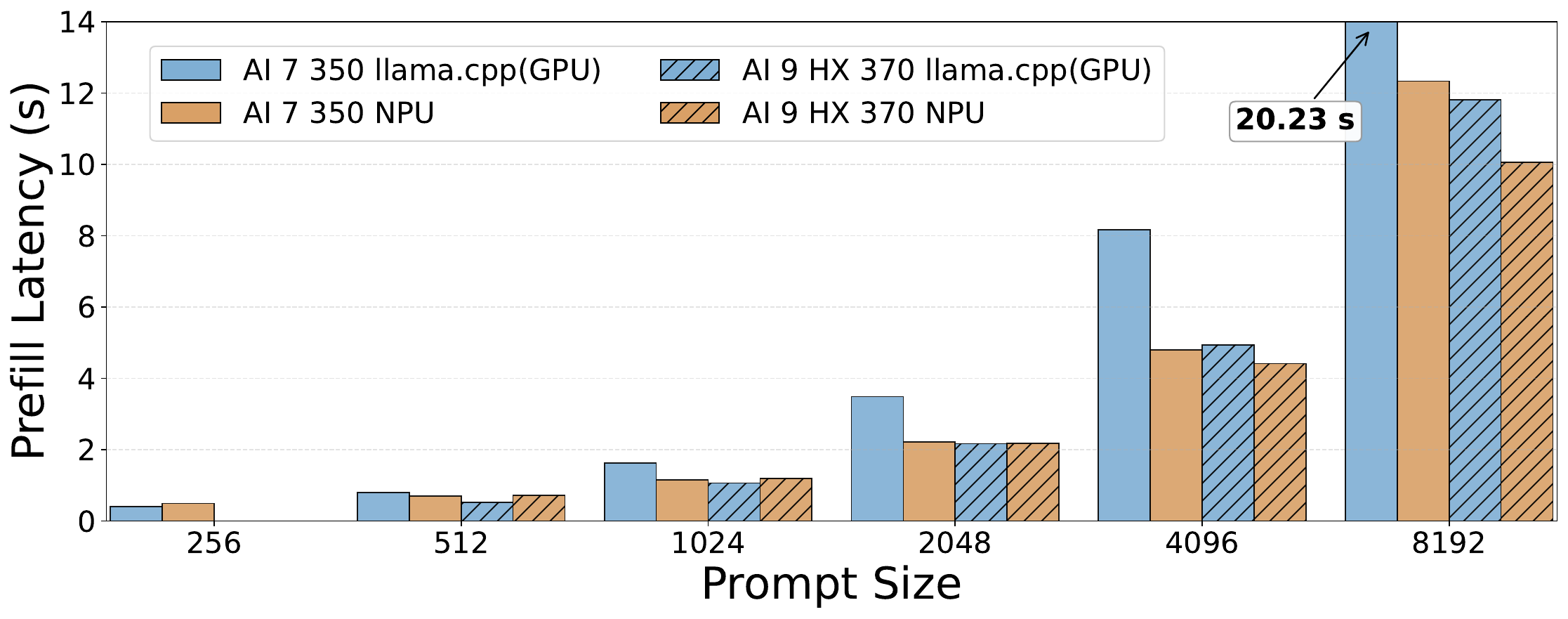}
    \vspace{-5pt}
    \caption{Qwen2.5-3B}
    \label{fig:e2eGemmaprefilTwo}
  \end{subfigure}

  \vspace{0.5em}

  \begin{subfigure}[b]{\linewidth}
    \centering
    \trimplot{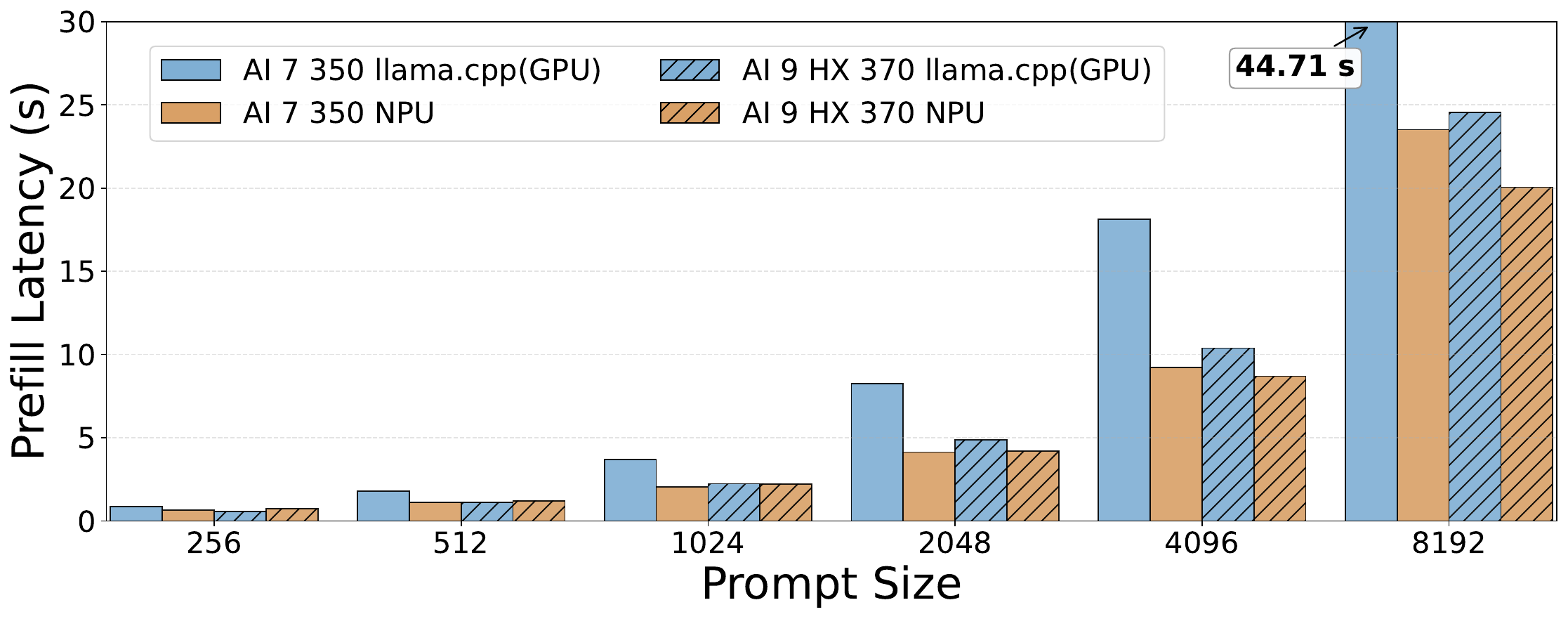}
    \vspace{-5pt}
    \caption{Llama3-8B}
    \label{fig:e2eLlamaPrefil}
  \end{subfigure}
  \caption{Comparison of W4A16 AWQ prefilling performance. We benchmark our NPU framework against the \texttt{llama.cpp} iGPU baseline. The $y$-axis represents execution latency while the $x$-axis shows prompt length.}
  \label{fig:e2eVisualizationprefilprefil}
\end{figure}

\begin{figure}[t!]
  \centering
   \includegraphics[width=\linewidth]{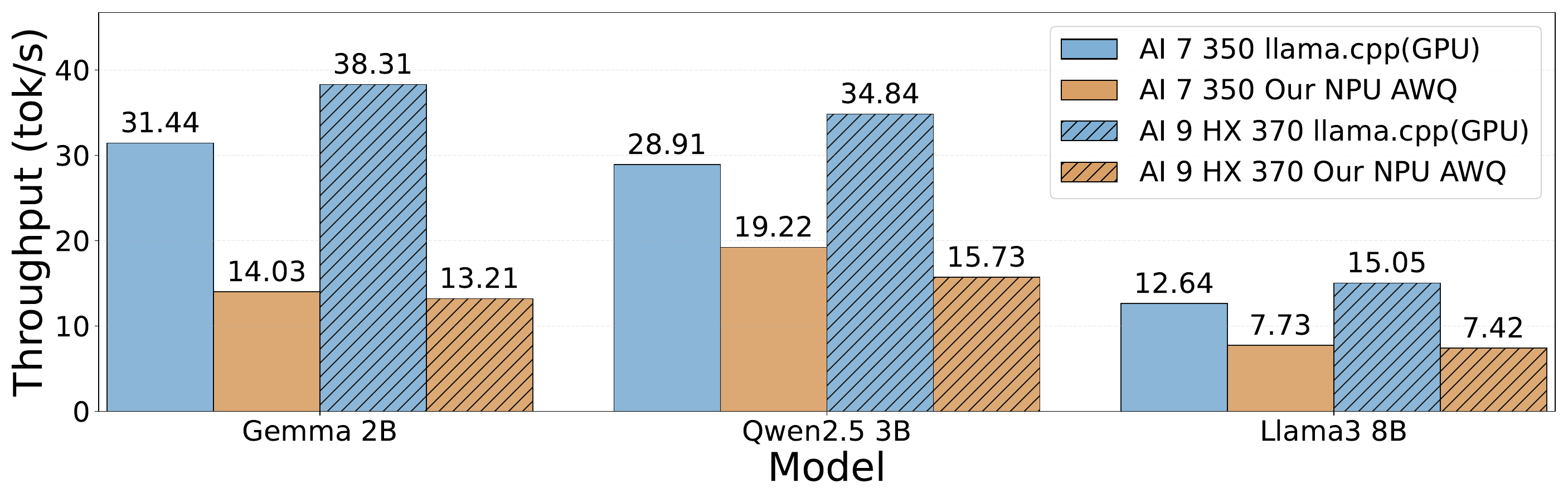}
   \caption{Comparison of W4A16 AWQ generation performance. We benchmark our NPU framework against the \texttt{llama.cpp} GPU baseline. The $y$-axis shows generation throughput in tokens/s.}
  \label{fig:e2eVisualizationgen}
\end{figure}

\begin{table*}[tb]
\setlength{\tabcolsep}{3pt}
\renewcommand{\arraystretch}{1.12}
\centering
\small
\caption{LLM prefill latency by prompt size on Ryzen AI 7 350 and Ryzen AI 9 HX 370. GPU and NPU report latency in milliseconds. Speedup is GPU latency divided by NPU latency.}
\label{tab:llm_prefill_latency_by_prompt}
\begin{tabular}{lcc | cc | c | cc | c}
\toprule
\multirow{2}{*}{\textbf{Model}} & \multirow{2}{*}{\textbf{Precision}} & \multirow{2}{*}{\textbf{Prompt}} & \multicolumn{3}{c|}{\textbf{AMD Ryzen AI 7 350}} & \multicolumn{3}{c}{\textbf{AMD Ryzen AI 9 HX 370}} \\
\cmidrule(lr){4-6}\cmidrule(lr){7-9}
& & & \textbf{GPU} & \textbf{NPU} & \textbf{Speedup} & \textbf{GPU} & \textbf{NPU} & \textbf{Speedup} \\
\midrule[\heavyrulewidth]
\multirow{12}{*}{\makecell[c]{Llama3\\8B}} & W4A16 & 256 & 869.7 ms & 650.8 ms & 1.34x & 569.3 ms & 722.2 ms & 0.79x \\
 & W4A16 & 512 & 1801.0 ms & 1101.5 ms & 1.64x & 1116.6 ms & 1210.1 ms & 0.92x \\
 & W4A16 & 1024 & 3700.7 ms & 2048.5 ms & 1.81x & 2240.7 ms & 2219.4 ms & 1.01x \\
 & W4A16 & 2048 & 8246.9 ms & 4129.9 ms & 2.00x & 4858.9 ms & 4194.9 ms & 1.16x \\
 & W4A16 & 4096 & 18141.9 ms & 9209.2 ms & 1.97x & 10380.7 ms & 8700.4 ms & 1.19x \\
 & W4A16 & 8192 & 44707.0 ms & 23506.2 ms & 1.90x & 24544.4 ms & 20043.8 ms & 1.22x \\
\cmidrule(lr){2-9}
 & W8A16 & 256 & 862.3 ms & 662.4 ms & 1.30x & 671.3 ms & 1045.7 ms & 0.64x \\
 & W8A16 & 512 & 1690.5 ms & 1099.2 ms & 1.54x & 1212.2 ms & 1654.8 ms & 0.73x \\
 & W8A16 & 1024 & 3545.0 ms & 2080.8 ms & 1.70x & 2341.1 ms & 3140.2 ms & 0.75x \\
 & W8A16 & 2048 & 7638.3 ms & 4284.1 ms & 1.78x & 4919.1 ms & 6629.1 ms & 0.74x \\
 & W8A16 & 4096 & 17613.2 ms & 9923.6 ms & 1.77x & 10291.2 ms & 15546.9 ms & 0.66x \\
 & W8A16 & 8192 & 44569.3 ms & 24510.9 ms & 1.82x & 23983.5 ms & 38539.4 ms & 0.62x \\
\cmidrule(lr){1-9}
\multirow{12}{*}{\makecell[c]{Gemma\\2B}} & W4A16 & 256 & 301.6 ms & 441.8 ms & 0.68x & 217.1 ms & 458.6 ms & 0.47x \\
 & W4A16 & 512 & 566.8 ms & 552.5 ms & 1.03x & 389.2 ms & 559.6 ms & 0.70x \\
 & W4A16 & 1024 & 1188.5 ms & 888.7 ms & 1.34x & 771.4 ms & 832.1 ms & 0.93x \\
 & W4A16 & 2048 & 2659.7 ms & 1566.1 ms & 1.70x & 1690.4 ms & 1503.4 ms & 1.12x \\
 & W4A16 & 4096 & 5638.8 ms & 3253.2 ms & 1.73x & 3841.5 ms & 3066.1 ms & 1.25x \\
 & W4A16 & 8192 & 13030.9 ms & 8343.6 ms & 1.56x & 9943.3 ms & 7097.7 ms & 1.40x \\
\cmidrule(lr){2-9}
 & W8A16 & 256 & 295.1 ms & 495.5 ms & 0.60x & 220.3 ms & 546.5 ms & 0.40x \\
 & W8A16 & 512 & 559.5 ms & 562.0 ms & 1.00x & 410.8 ms & 808.8 ms & 0.51x \\
 & W8A16 & 1024 & 1147.6 ms & 952.9 ms & 1.20x & 860.9 ms & 1393.8 ms & 0.62x \\
 & W8A16 & 2048 & 2517.3 ms & 1870.6 ms & 1.35x & 1719.1 ms & 2859.7 ms & 0.60x \\
 & W8A16 & 4096 & 5114.9 ms & 4187.0 ms & 1.22x & 3296.2 ms & 6489.7 ms & 0.51x \\
 & W8A16 & 8192 & 12029.8 ms & 10219.0 ms & 1.18x & 7624.7 ms & 15990.4 ms & 0.48x \\
\cmidrule(lr){1-9}
\multirow{12}{*}{\makecell[c]{Qwen2.5\\3B}} & W4A16 & 256 & 402.5 ms & 490.1 ms & 0.82x & 303.0 ms & 598.2 ms & 0.51x \\
 & W4A16 & 512 & 790.0 ms & 696.1 ms & 1.13x & 518.2 ms & 714.6 ms & 0.73x \\
 & W4A16 & 1024 & 1629.6 ms & 1155.3 ms & 1.41x & 1064.0 ms & 1190.6 ms & 0.89x \\
 & W4A16 & 2048 & 3482.6 ms & 2216.7 ms & 1.57x & 2163.0 ms & 2173.9 ms & 0.99x \\
 & W4A16 & 4096 & 8158.5 ms & 4800.3 ms & 1.70x & 4936.0 ms & 4409.2 ms & 1.12x \\
 & W4A16 & 8192 & 20231.4 ms & 12334.4 ms & 1.64x & 11804.0 ms & 10060.5 ms & 1.17x \\
\cmidrule(lr){2-9}
 & W8A16 & 256 & 386.4 ms & 449.8 ms & 0.86x & 399.0 ms & 518.8 ms & 0.77x \\
 & W8A16 & 512 & 758.3 ms & 702.8 ms & 1.08x & 585.4 ms & 786.6 ms & 0.74x \\
 & W8A16 & 1024 & 1563.4 ms & 1287.4 ms & 1.21x & 1098.9 ms & 1399.0 ms & 0.79x \\
 & W8A16 & 2048 & 3310.3 ms & 2663.6 ms & 1.24x & 2249.5 ms & 2870.5 ms & 0.78x \\
 & W8A16 & 4096 & 7691.8 ms & 6091.1 ms & 1.26x & 4830.4 ms & 6604.9 ms & 0.73x \\
 & W8A16 & 8192 & 19749.1 ms & 14959.2 ms & 1.32x & 10599.8 ms & 16249.8 ms & 0.65x \\
\bottomrule
\end{tabular}
\end{table*}

With our custom end-to-end framework setup introduced in Section 4.6, we evaluate the end-to-end performance. Note that the Ryzen AI 7 350 provides a more informative comparison for isolating NPU-side benefits because its iGPU baseline is substantially smaller, while the HX 370 has a stronger iGPU that can partially mask NPU-side gains in end-to-end comparisons. The end-to-end performance of the evaluated LLMs is shown in Figure~\ref{fig:e2eVisualizationprefilprefil} and Figure~\ref{fig:e2eVisualizationgen}. Full request latency can be estimated as \(L = l_p + n \cdot l_g\), where \(l_p\) is prefilling latency, \(l_g\) is per-token generation latency (the inverse of tok/s), and \(n\) is the number of generated output tokens. Figure~\ref{fig:e2eVisualizationprefilprefil} reports prefilling latency across prompt lengths, while Figure~\ref{fig:e2eVisualizationgen} reports generation throughput. We further break down prefilling latency for both $W4A16$ and $W8A16$ in Table~\ref{tab:llm_prefill_latency_by_prompt}.

A clear pattern emerges for $W4A16$. On the Ryzen AI 7 350, our NPU implementation matches or outperforms the \texttt{llama.cpp} iGPU baseline for medium and long prompts, reaching up to 2.0$\times$ speedup for Llama3-8B and 1.7$\times$ for Gemma-2B and Qwen2.5-3B. For short prompts, the iGPU generally delivers lower latency because fixed NPU overheads from kernel dispatch and fabric reconfiguration dominate small matrix shapes. As prompt length increases, these costs are amortized and the NPU advantage becomes more visible.

The $W8A16$ results show a weaker trend. On the Ryzen AI 7 350, $W8A16$ still provides speedups over the iGPU for many prompt lengths, but the gains are smaller than $W4A16$; for example, Llama3-8B reaches 1.82$\times$ at 8192 tokens compared with 1.90$\times$ for $W4A16$, while Gemma-2B and Qwen2.5-3B show more modest speedups. On the Ryzen AI 9 HX 370, however, the $W8A16$ NPU path is slower than the iGPU baseline across the measured models and prompt lengths, consistent with the kernel results in the previous section, where the less compact $W8A16$ kernels have lower NPU throughput on the Ryzen AI 9 HX 370. At the same time, the \texttt{llama.cpp} iGPU baseline does not slow down proportionally when moving from \texttt{Q4\_K} to \texttt{Q8\_0}; in several cases, \texttt{Q8\_0} is similar to or slightly faster than \texttt{Q4\_K}. We attribute this to \texttt{llama.cpp}'s block-wise fused dequantization: although \texttt{Q8\_0} moves more weight data, its dequantization path is simpler than the more complex 4-bit K-quant layout, reducing some of the per-block unpacking and scaling work inside the fused GPU kernel. 

Comparing the Ryzen AI 7 350 and Ryzen AI 9 HX 370 reveals an important system-level tradeoff. Although the 350 shows stronger standalone $W4A16$ NPU kernel performance, its end-to-end prefilling latency remains close to the HX 370. We attribute this to the 350's weaker iGPU, which bottlenecks FlashAttention-2 and offsets its NPU advantage. By contrast, the HX 370 benefits from a 16-CU iGPU that compensates for its weaker NPU kernel performance, leading to similar final prefilling latency across the two devices for $W4A16$.

Token generation shows the opposite trend in Figure~\ref{fig:e2eVisualizationgen}. Here, the NPU underperforms the iGPU because generation is fundamentally latency-bound: GEMV operators execute in hundreds of microseconds, but NPU driver and reconfiguration overheads remain in the millisecond range. We also observe mapping inefficiencies for models such as Qwen2.5-3B, whose MLP hidden dimension is not a multiple of our minimum tile-size. The resulting padding increases over-computation and causes the NPU to fall behind the iGPU in latency.

\subsection{Power Measurements}

\begin{figure}[t!]
  \centering
   \includegraphics[width=\linewidth]{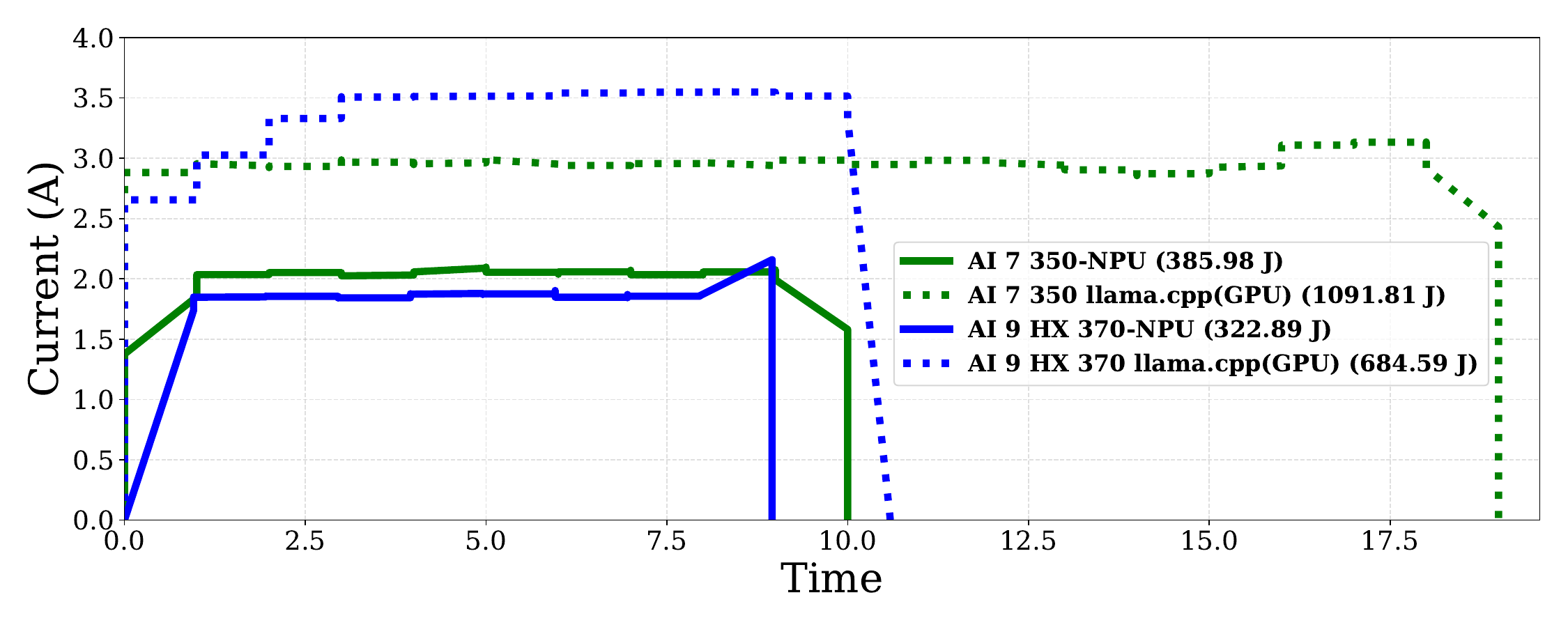}
   \caption{Current traces for Llama3-8B prefilling with a 4096-token prompt on the Ryzen AI 7 350 and Ryzen AI 9 HX 370.}
  \label{fig:e2ePower}
\end{figure}

\begin{figure*}[t!]
  \centering
  \begin{subfigure}[t]{0.49\textwidth}
    \centering
    \includegraphics[width=\linewidth]{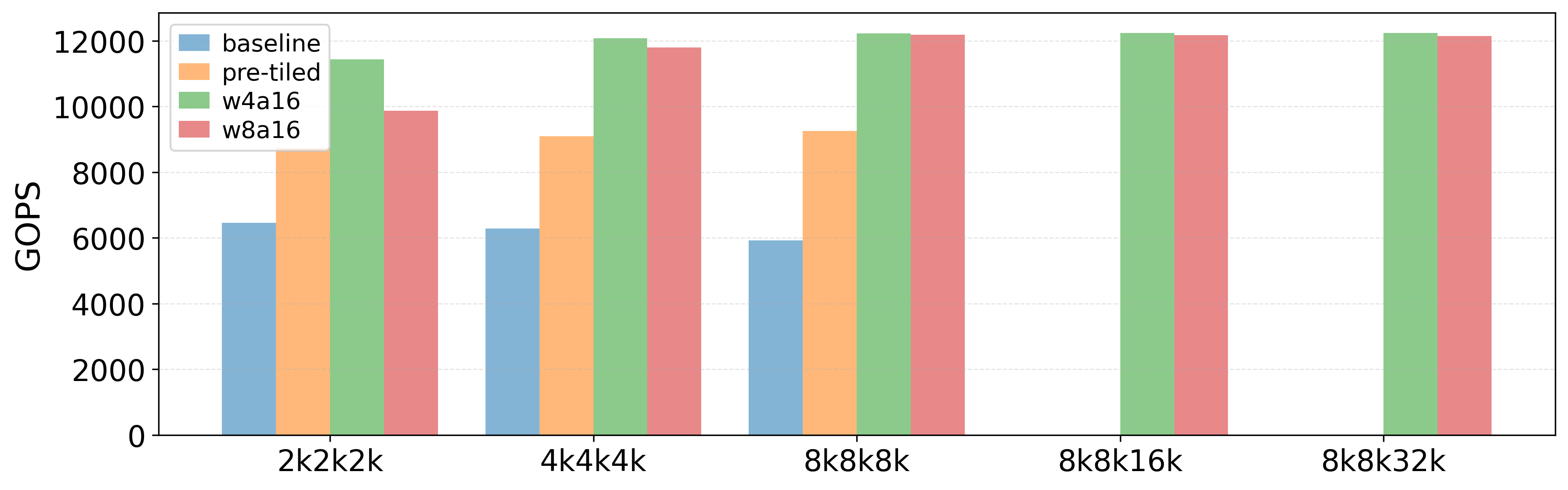}
    \caption{AMD Ryzen AI 7 350: Krackan Point}
    \label{fig:GEMM_krackan_ablation}
  \end{subfigure}
  \begin{subfigure}[t]{0.49\textwidth}
    \centering
    \includegraphics[width=\linewidth]{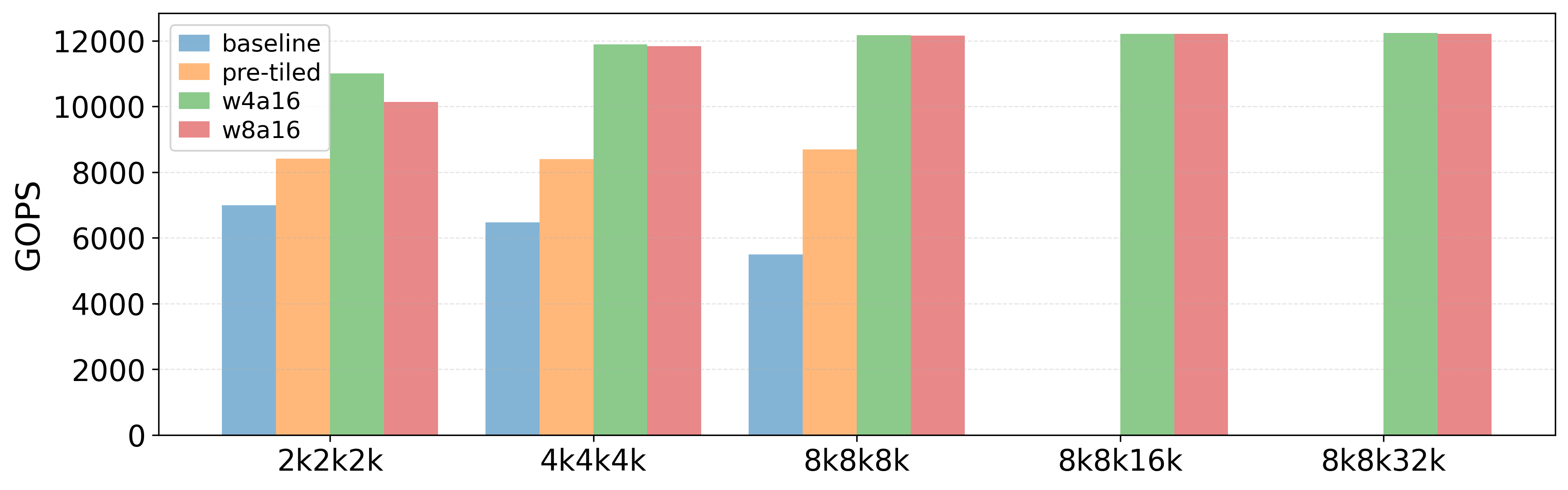}
    \caption{AMD Ryzen AI 9 HX 370: Strix Point}
    \label{fig:GEMM_strix_ablation}
  \end{subfigure}\hfill
  \caption{Performance of NPU kernels across different GEMM sizes and optimizations for ablation study. Baseline/pre-tiled do not support 8k×8k×16k and 8k×8k×32k without our interleaved pre-tiling, therefore left empty.}
  \label{fig:GEMM_ablation}
\end{figure*}

\begin{figure*}[t!]
  \centering
  \begin{subfigure}[t]{0.49\textwidth}
    \centering
    \includegraphics[width=\linewidth]{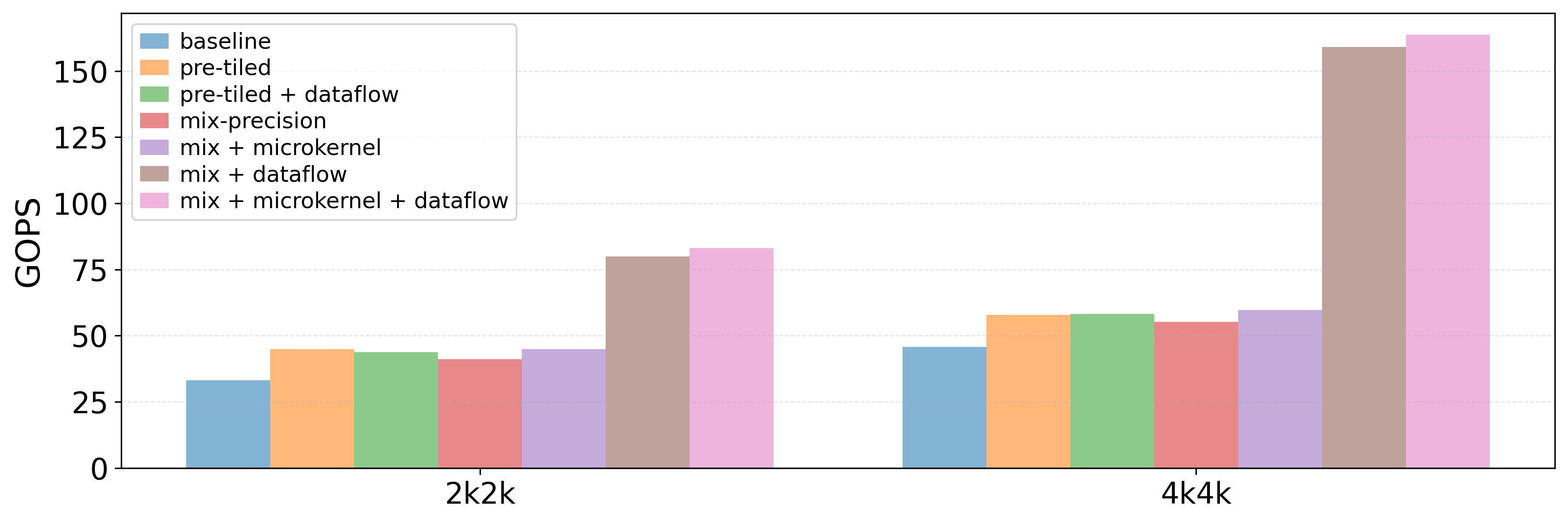}
    \caption{AMD Ryzen AI 7 350: Krackan Point}
    \label{fig:GEMV_krackan_ablation}
  \end{subfigure}
  \begin{subfigure}[t]{0.49\textwidth}
    \centering
    \includegraphics[width=\linewidth]{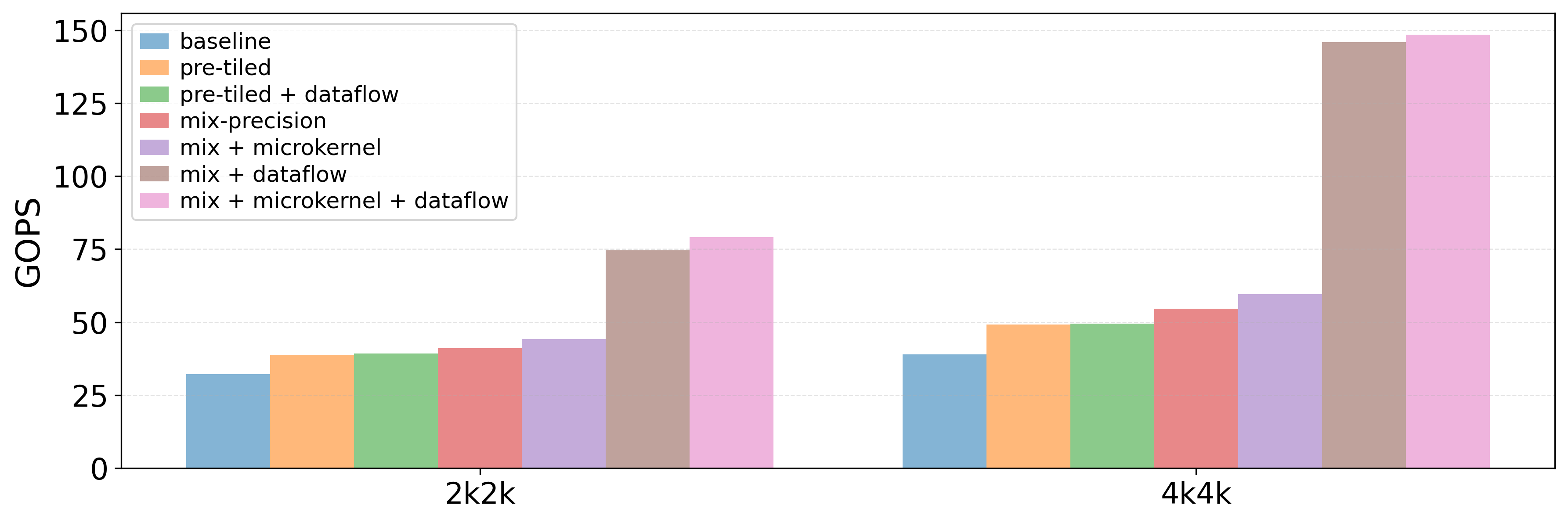}
    \caption{AMD Ryzen AI 9 HX 370: Strix Point}
    \label{fig:GEMV_strix_ablation}
  \end{subfigure}\hfill
  \caption{Performance of NPU kernels across different GEMV sizes and optimizations for ablation study.}
  \label{fig:GEMV_ablation}
\end{figure*}

We measure platform current during LLM inference using a lab bench power supply, and use the resulting electrical traces to compare system energy and instantaneous current draw across execution modes~\cite{owon_spe_series_dc_power_supply_2026}. The resulting traces are shown in Figure~\ref{fig:e2ePower}. The figure reports Llama3-8B prefilling with a 4096-token prompt; the same qualitative trend also holds across the other evaluated models and prompt lengths, which are omitted here for brevity. Although the NPU path offloads only GEMMs to the NPU while leaving the remaining nonlinear operations on the GPU, it still substantially reduces total energy. On the Ryzen AI 7 350, the NPU path lowers energy consumption by 64.6\%, largely because it completes the workload both faster and more efficiently. On the Ryzen AI 9 HX 370, where execution times are closer, the NPU path still reduces total energy by 52.8\% due to its higher efficiency. In both cases, moving prefilling to the NPU also lowers the instantaneous current draw throughout most of execution.

\subsection{Comparison with Other NPU Works}
\begin{table}[h]
  \centering
  \resizebox{\columnwidth}{!}{%
  \begin{tabular}{llccr}
    \toprule
    \textbf{Model} & \textbf{Prior Work} & \textbf{Ref. Speed} & \textbf{Ours} & \textbf{Speedup} \\
    & & (tok/s) & (tok/s) & \\
    \midrule
    \textbf{Llama-3-8B} & FastFlowLM~\cite{fastflowlm_github_2026} & 345.9 & \textbf{501.9} & \textbf{1.45$\times$} \\
    \textbf{Qwen2.5-3B} & Scaling NPU~\cite{scaling_llm_mobile_2025} & 550.0 & \textbf{890.4} & \textbf{1.61$\times$} \\
    \bottomrule
  \end{tabular}%
  }
    \caption{Comparison of Prefilling Throughput (tokens/s) at a context length of 1024 tokens. Our end-to-end flow using our fused NPU kernel demonstrates superior performance compared to recent NPU-driven Edge LLM inference works.}
  \label{tab:sota_comparison}
  \vspace{-10pt}
\end{table}

In Table~\ref{tab:sota_comparison}, we compare our results against recent NPU-driven inference frameworks. Direct apples-to-apples comparison is difficult because platforms differ in hardware, power limits, and quantization schemes. To reduce this mismatch, we choose baselines aligned with our target model families and 4-bit setting, and report results on the Ryzen AI 7 350, which performed best in our evaluation. The results show that our fused mixed-precision XDNA2 framework is highly competitive within the current edge-AI landscape.

\subsection{Ablation Study}

Figure~\ref{fig:GEMM_ablation} and Figure~\ref{fig:GEMV_ablation} isolate the contribution of each optimization. We also use this ablation to quantify the benefit of leveraging XDNA2 reconfigurability. Since we cannot physically disable the reconfigurability of the AIE array, we define the baseline as the closest fixed-kernel counterpart: standard full-precision (BF16) GEMM/GEMV kernels with the original weight layout and data movement pattern, without the workload-specific optimizations introduced in this paper. Each added optimization represents one way TileFuse exploits reconfigurability: pre-tiling adapts the memory layout and DMA access pattern, mixed-precision kernels customize the compute-core datapath for fused unpacking and dequantization, and the GEMV microkernel/dataflow optimizations change the per-core work granularity and array-level weight distribution.

For GEMM, the fixed baseline does not support output dimensions above 8K, whereas our mixed-precision kernels with the interleaved column-major layout scale to 32K without a noticeable performance drop. The baseline also degrades slightly as GEMM size increases, consistent with inefficient weight streaming from DRAM under large strides. Applying pre-tiling alone improves the full-precision baseline by up to 50\%, showing the benefit of adapting the memory layout and DMA streaming pattern to the AIE array. Building on this layout, mixed-precision kernels outperform the fixed baseline by roughly 50\% even at smaller sizes and by up to 121.6\% at larger sizes, demonstrating the additional gain from customizing the compute-core datapath to directly consume quantized weights.

Between the two mixed-precision formats, the faster choice depends on the balance between weight-streaming bandwidth and dequantization overhead. $W4A16$ transfers fewer weight bytes, but it requires packed INT4 unpacking and asymmetric zero-point handling. In contrast, $W8A16$ transfers more data, but uses symmetric INT8 weights with a simpler dequantization path. For GEMMs with smaller $M$, each streamed weight tile is reused across fewer activation rows, resulting in lower arithmetic intensity and making weight-traffic reduction more important. As $M$ increases, weight reuse and arithmetic intensity improve, so the bandwidth advantage of $W4A16$ becomes less dominant and dequantization overhead can have a larger impact on the relative performance of $W4A16$ and $W8A16$. Platform-specific memory behavior can also affect the relative performance, as discussed in Section~6.3.

For GEMV, pre-tiling alone improves the full-precision baseline by more than 20\%, while fused $W4A16$ further reduces memory pressure by transmitting INT4 rather than BF16. However, its benefit remains modest at roughly 20\%--40\% without further optimization because unpacking and dequantization introduce extra compute. The microkernel and dataflow optimizations primarily target computation throughput, so they offer little benefit on the original memory-bound baseline. Once mixed precision shifts GEMV toward a more compute-bound regime, however, they become effective. Combining mixed precision with both the optimized microkernel and the 2D weight-distribution dataflow yields up to 281\% improvement over the fixed baseline, showing the benefit of using XDNA2 reconfigurability not only for the compute-core datapath, but also for array-level data movement, as shown in Figure~\ref{fig:GEMV_ablation}a.

\section{Discussion and Limitations}

\subsection{Generalizability Across Platforms}
TileFuse generalizes in two important ways. First, on AMD Ryzen AI systems, it enables widely used AWQ-style models to run natively on the NPU, rather than requiring the model to be reshaped around an NPU-specific quantization format. This makes XDNA2 a more practical target for off-the-shelf edge LLM deployment and, because XDNA2 is deployed across current-generation Ryzen AI SoCs, gives the broader Ryzen AI family a reusable path toward more accessible NPU acceleration. Second, while our specific microkernels and dataflow optimizations depend on XDNA2’s reconfigurable spatial array and close-to-metal control, the underlying pre-tiling idea is more portable. Offline reorganization of AWQ weights to match hardware consumption patterns can benefit other GPUs and NPUs as well, although the exact layout must be adapted to the architecture and programming model of the target platform.

\subsection{A Hybrid Solution}

XDNA2 offers substantial reconfigurability, enabling kernel- and dataflow-level optimization across diverse GEMM shapes, but this flexibility comes with non-trivial runtime overhead from kernel dispatch and fabric reconfiguration. During prefilling, where GEMM operators often run for tens of milliseconds, this overhead is largely amortized; during token generation, however, batch-1 GEMV operators often execute in only hundreds of microseconds, making runtime overhead a dominant cost and explaining why the NPU underperforms the iGPU for token generation despite TileFuse’s kernel-level optimizations. 

These results point to a practical hybrid deployment strategy for interactive edge inference: the NPU is best suited for long-context prefilling, where large GEMM workloads can exploit its reconfigurable spatial array efficiently, while the iGPU remains a better target for latency-sensitive token generation. Our current implementation already moves in this direction by offloading attention and miscellaneous operations to the iGPU, and because AMD Ryzen SoCs use unified memory across the CPU, iGPU, and NPU, such hybrid execution does not introduce significant coherence or data-movement overhead. Extending this approach, a system that performs prefilling on the NPU and token generation on the iGPU could combine the prefilling gains in Figure~\ref{fig:e2eVisualizationprefilprefil} with the higher generation throughput in Figure~\ref{fig:e2eVisualizationgen}.

\subsection{Strix Point vs. Krackan Point}

Krackan Point machines (Ryzen AI 7 350) are positioned lower than Strix Point machines (Ryzen AI 9 HX 370). However, our kernel, end-to-end, and ablation results show that the Ryzen AI 7 350 achieves higher NPU performance than the Ryzen AI 9 HX 370 in several cases, especially after pre-tiling is applied, as shown in Figure~\ref{fig:kernel_comparison}, Figure~\ref{fig:GEMM_ablation} and Figure~\ref{fig:GEMV_ablation}.

This behavior is not fully explained by public platform specifications. Both processors support similar memory configurations, so the difference is unlikely to come from nominal peak DRAM bandwidth alone. Instead, the results suggest that the two platforms may differ in effective NPU-side streaming bandwidth, memory-path overhead, power/thermal behavior, or runtime scheduling overhead. AMD does not publicly document the detailed NPU memory-path differences needed to isolate the exact cause, so we treat this as an empirical platform-level observation.

The format-dependent behavior in Figure~\ref{fig:kernel_comparison} is consistent with this interpretation. Although $W4A16$ transfers fewer weight bytes than $W8A16$, it also has a more complex dequantization path because packed INT4 weights require unpacking and asymmetric zero-point handling. By contrast, $W8A16$ transfers more data but uses symmetric INT8 weights with simpler dequantization. Therefore, $W4A16$ is not guaranteed to always outperform $W8A16$: when the platform can stream the larger INT8 weight tiles efficiently, the lower dequantization overhead of $W8A16$ can make it competitive or even faster. This appears in several Ryzen AI 7 350 results. On the Ryzen AI 9 HX 370, however, the larger $W8A16$ data movement appears to be more costly, causing $W8A16$ to fall behind $W4A16$ despite its simpler arithmetic.

Overall, these results suggest that the observed performance difference between Strix Point and Krackan Point comes from the interaction between data layout, effective NPU-side streaming behavior, dequantization overhead, and platform-specific runtime behavior, rather than from nominal DRAM bandwidth alone. This also explains why pre-tiling is particularly important: by improving memory contiguity and reducing inefficient DMA access patterns, it allows platforms with better effective streaming behavior to benefit more from the optimized layout.

\section{Conclusion}
We presented \textit{TileFuse}, a close-to-metal mixed-precision kernel library for AMD XDNA2 NPUs that brings practical off-the-shelf LLM quantization formats such as AWQ-style $W4A16$ directly onto the NPU. 

Our results show that TileFuse supports GEMM dimensions up to 32K, compared with the baseline's 8K limit, improves kernel-level performance by up to 121.6\% for GEMM and 281\% for GEMV, and achieves up to $2.0\times$ lower prefilling latency with more than 64.6\% lower energy than a tuned iGPU baseline on Ryzen AI systems. More broadly, these results show that NPUs become substantially more useful for edge LLM deployment as we adapt hardware to the quantization formats users already rely on.

\section*{Acknowledgment}
We thank Chengyue Wang from UCLA and Joe Melber from AMD for their insightful feedback on this work. 
This work is supported in part by the AMD Center of Excellence at UIUC. 
We also thank AMD for providing the Ryzen AI platforms used in this work.


\bibliographystyle{ACM-Reference-Format}
\bibliography{reference}










\end{document}